\documentclass[aip,pof,reprint]{revtex4-1}

\usepackage{amsmath}
\usepackage{graphicx}
\usepackage{color}

\usepackage{natbib}

\newcommand\omegab{\boldsymbol{\omega}}
\newcommand\taub{\boldsymbol{\tau}}

\providecommand\upi{\pi}

\newcommand{\td}[2] {\frac{d #1}{d #2}}

\newcommand{\pd}[2] {\frac{\partial #1}{\partial #2}}

\renewcommand{\vec}[1]{{\bf#1}}

\renewcommand{\div}{\vec{\nabla \cdot}}

\newcommand{\grad}{\vec \nabla }

\newcommand{\T}{\theta}

\begin{document}

\title{The effect of particle density in turbulent channel flow laden with finite size particles in semi-dilute conditions} 



\author{W.~Fornari}
\affiliation{SeRC and Linn\'e FLOW Centre, KTH Mechanics, SE-100 44 Stockholm, Sweden}

\author{A.~Formenti}
\affiliation{Department of Industrial Engineering, University of Padova, Via Venezia 1, 35131, Padova, Italy}

\author{F.~Picano}
\affiliation{Department of Industrial Engineering, University of Padova, Via Venezia 1, 35131, Padova, Italy}

\author{L.~Brandt}
\affiliation{SeRC and Linn\'e FLOW Centre, KTH Mechanics, SE-100 44 Stockholm, Sweden}


\date{\today}

\begin{abstract}

We study the effect of varying the mass and volume fraction of a suspension of rigid spheres dispersed 
in a turbulent channel flow. We performed several Direct Numerical Simulations using an Immersed Boundary Method for 
finite-size particles changing the solid to fluid density ratio $R$, the mass fraction $\chi$ and the volume fraction $\phi$. 
We find that varying the density ratio $R$ between 1 and 10 at constant volume fraction does not alter the flow statistics 
as much as when varying the volume fraction $\phi$ at constant $R$ and at constant mass fraction. 
Interestingly, the increase in overall drag found when varying the volume fraction is considerably 
higher than that obtained for increasing density ratios at same volume fraction.
The main effect at 
density ratios $R$ of the order of 10 is a strong shear-induced migration towards the centerline of the channel. When the 
density ratio $R$ is further increased up to 1000 the particle dynamics decouple from that of the fluid. The solid phase behaves as a dense gas and 
the fluid and solid phase statistics drastically change. In this regime, the collision 
rate is high and dominated by the normal relative velocity among particles.

\end{abstract}

\pacs{}

\maketitle 


%
%

%




\section{Introduction}

The transport of particles in flows is relevant to many industrial applications and environmental processes. 
Examples include sediment transport in rivers, avalanches and pyroclastic flows, as well as many oil industry and 
pharmaceutical processes. Often the flow regime encountered in such applications is turbulent due to the high flow 
rates and it can be substantially affected by the presence of the solid phase. Depending on the features of both 
fluid and solid phases, many different scenarios can be observed and the understanding of such flows is still incomplete.\\
The rheological properties of these suspensions have mainly been studied in the viscous Stokesian regime and in the low speed 
laminar regime. Even limiting our attention to monodisperse rigid neutrally buoyant spheres suspended in Newtonian 
liquids, we find interesting rheological behaviors such as shear thinning or thickening, jamming at high volume fractions, 
and the generation of high effective viscosities and normal stress differences \cite{stickel2005,morris2009,wagner2009}.
It is known that the effective viscosity of a suspension $\mu_e$ changes with respect to that of the pure fluid $\mu$ 
due to the modification of the response of the complex fluid to the local deformation rate\cite{guazz2011}. In the dilute 
regime, an expression for the effective viscosity $\mu_e$ with the solid volume fraction $\phi$ has first been proposed by 
Einstein\cite{einstein1906,einstein1911} and then corrected by Batchelor\cite{batchelor1970} and Batchelor and 
Green\cite{batchelor1972}. As the volume fraction increases, the mutual interactions among particles become more important 
and the effective viscosity increases until the system jams\cite{sierou2002}. At high volume fractions, the variation of the 
effective viscosity $\mu_e$ is described exclusively by semi-empirical laws such as those by Eiler and Krigher \& Dougherty
\cite{stickel2005} that also capture the observed divergence at the maximum packing limit, \cite{boyer2011} $\phi_m=0.58-0.62$. 
In laminar flows, shear-thickening or normal stress differences occur due to inertial effects at the 
particle scale. Indeed, when the particle Reynolds number $Re_a$ is non negligible the symmetry of the particle pair trajectories 
is broken and the microstructure becomes anisotropic, leading to macroscopical behaviors such as shear-thickening \cite{Morrispof08,picano2013,Morris2014}.
Finally, in the highly inertial regime 
the effective viscosity $\mu_e$ increases 
linearly with shear rate due to augmented particle collisions \cite{bagnold1954}.

Another important feature observed in viscous flows is shear-induced migration. When considering a pressure-driven 
Poiseuille flow, either in a tube or in a channel, the particles irreversibly migrate toward the centerline, i.e. from 
high to low shear rate regions\cite{guazz2011,koh1994}. Interestingly when inertial effects become important, a different kind of 
migration occurs as the particles tend to move radially away from both the centerline and the walls, toward an intermediate equilibrium position. 
This type of migration was first observed in a tube \cite{guazz2011,segre1962} and was named tubular pinch. It is mechanistically 
unrelated to the rheological properties of the flow and results from the fluid-particle interaction within the conduit. The case of 
the laminar square duct flow has also been studied to identify the particle equilibrium 
positions \cite{chun2006,abbas2014}. It was found that finite-size particles migrate toward the corners or to the center of edges  
depending on the bulk Reynolds number. At high Reynolds numbers (but still in the laminar regime), some particles were also found 
in an inner region near the center of the duct.

Typically, as the Reynolds number is increased inertial effects become important and the unladen flow undergoes a transition from laminar 
to turbulent conditions. The presence of the solid phase may alter this process by either increasing or reducing the 
critical Reynolds number above which the transition to the turbulent regime occurs. 
The case of a dense suspension of particles in a 
pipe flow has been studied experimentally \cite{matas2003} and numerically \cite{yu2013}. It has been suggested that transition 
depends upon the pipe to particle diameter ratios and the volume fraction. 
For larger particles, transition shows a non-monotonic behavior that cannot be solely explained in terms of an increase of the effective viscosity. 
For smaller neutrally-buoyant particles instead, the critical Reynolds number increases monotonically with the solid volume fraction due to the raise 
in effective viscosity.

The transition in dilute suspensions of finite-size 
particles in plane channels has been studied by Lashgari et al. \cite{lashg2015} and Loisel et al. \cite{loisel2013}. It has been shown that 
the critical Reynolds number above which turbulence is sustained, is reduced. At fixed Reynolds number and solid volume fraction, the 
initial arrangement of particles is important to trigger the transition. 
Lashgari et al. \cite{lashgari2014} also investigated numerically a channel flow laden with solid spherical particles at higher volume fractions and 
for a wide range of Reynolds numbers. These authors identified three different regimes for different values of the solid volume fraction $\phi$ and the 
Reynolds number $Re$. In each regime (laminar, 
turbulent and inertial shear-thickening), the flow is dominated by different components of the total stress (viscous, 
turbulent or particle stresses respectively).

Regarding the fully turbulent regime, most of the previous studies have focused on dilute or very dilute suspensions of particles smaller than the 
hydrodynamic scales and heavier than the fluid. In the one-way coupling regime \cite{balach-rev2010} (i.e. when the solid phase has a negligible 
effect on the fluid phase) and limiting our attention to wall-bounded flows, it has been shown that particles migrate from regions of 
high to low turbulence intensities \cite{reeks1983}. This phenomenon is known as turbophoresis and it has been shown to be stronger 
when the turbulent near-wall characteristic time and the particle inertial time scale are similar \cite{soldati2009}. Small-scale clustering 
has also been observed in this kind of inhomogeneous flows \cite{sardina2012}, leading together with turbophoresis to the formation of 
streaky particle patterns \cite{sardina2011}. In the two-way coupling regime (i.e. when the mass density ratios are high and the 
back-reaction of the dispersed phase on the fluid cannot be neglected) the solid phase has been shown to reduce the turbulent near-wall 
fluctuations increasing their anisotropy \cite{kulick1994} and eventually reducing the total drag \cite{zhao2010}.

When the suspensions are dense it is of fundamental importance to consider particle-particle interactions and collisions. Indeed, the chaotic 
dynamics of the fluid phase affects the rheological properties of the suspension, especially at high Reynolds numbers. This is 
known as a four-way coupling regime.  
Increasing the particle size directly affects the turbulent structures at smaller and comparable scales \cite{naso2010} thereby modulating 
the turbulent field. In a turbulent channel flow it has been reported that finite-size particles larger than the dissipative length scale 
increase the turbulent intensities and the Reynolds stresses \cite{pan1996}. Particles are also found to preferentially accumulate in the 
near-wall low-speed streaks \cite{pan1996}. This has also been observed in open channel flows laden with heavy finite-size particles. In 
this case the flow structures are found to have a smaller streamwise velocity \cite{kida2013,kida2014}.

Concerning turbulent channel flows of neutrally buoyant particles, recent studies with $\phi \simeq 7\%$ report that due to the attenuation 
of the large-scale streamwise vortices, the fluid streamwise velocity fluctuation is reduced. When the particles are heavier than the carrier 
fluid and therefore sediment, the bottom wall acts as a rough boundary which makes the particles resuspend \cite{shao2012}. 
Recent simulations from our group have shown that the 
overall drag increases as the volume fraction is increased from $\phi=0\%$ up to $20\%$. This trend cannot be solely explained in terms 
of the increase of the suspension effective viscosity. It is instead found that as particle volume fraction increases, the velocity fluctuation 
intensities and the Reynolds shear stresses decrease while there is a significant increase of the particle induced stresses. The latter, in turn, 
lead to a higher overall drag \cite{picano2015}.

As noted by Prosperetti \cite{prosp2015}, however, results obtained for solid to fluid density ratios $R=\rho_p/\rho_f=1$ cannot be easily extrapolated 
to other cases (e.g. when $R > 1$). In the present study we therefore investigate numerically the effects of varying the density ratio $R$ of 
the suspended phase and consequently the mass fraction $\chi$ for different volume fractions. The aim is to understand separately  the 
effects of excluded volume and (particle and fluid) inertia on the statistical observables of both phases. To isolate the effects of different 
density ratios $R$ on the macroscopical behavior of the suspension, we consider an ideal situation where the effect of gravity is neglected, 
leaving its analysis to future studies.

We consider a turbulent channel flow laden with rigid spheres of radius $a=h/18$ where $h$ is the half-channel height (see Picano et al. 
\cite{picano2015}). Direct numerical simulations (DNS) fully describing the solid phase dynamics via an immersed boundary method (IBM) are 
performed as in Lucci et al. \cite{lucci2010} and Kidanemariam et al. \cite{kida2013} among others. First, cases at fixed mass fractions $\chi = 0.2$ 
are examined and compared to cases with constant volume fraction $\phi=5\%$ 
and density ratios $R$ ranging from $1$ to $10$. It is observed that the influence of the density ratio $R$ on the statistics of both phases 
is less important than that of an increasing volume fraction $\phi$. The main effects at density ratio $R \sim 10$ are shear-induced migration 
towards the centerline of the channel and slight reduction of the fluid velocity fluctuations in the log-layer. 
The results drastically change when further increasing $R$ (up to $\sim 1000$). It is found that for sufficiently high $R$ ($\gtrsim 100$), the solid phase 
behaves as a dense gas uncorrelated to the details of the carrier fluid flow.

\section{Methodology}
\subsection{Numerical method}

Different methods have been proposed in the last years to perform Direct Numerical Simulations of multiphase flows.
In the present study, simulations have been performed using the algorithm originally developed by Breugem\cite{breugem2012} that fully
describes the coupling between the solid and fluid phases. The Eulerian fluid phase is evolved according to the incompressible
Navier-Stokes equations,
\begin{equation}
\label{div_f}
\div \vec u_f = 0
\end{equation}
\begin{equation}
\label{NS_f}
\pd{\vec u_f}{t} + \vec u_f \cdot \grad \vec u_f = -\frac{1}{\rho_f}\grad p + \nu \grad^2 \vec u_f + \vec f
\end{equation}
where $\vec u_f$, $\rho_f$ and $\nu=\mu/\rho_f$ are the fluid velocity, density and kinematic viscosity respectively ($\mu$ is
the dynamic viscosity), while $p$ and $\vec f$ are the pressure and a generic force field (used to model the presence of particles). 
The particles centroid linear and angular velocities, $\vec u_p$ and $\vec \omegab_p$ are instead governed by the Newton-Euler Lagrangian 
equations,
\begin{align}
\label{lin-vel}
\rho_p V_p \td{\vec u_p}{t} &= \rho_f \oint_{\partial \mathcal{V}_p}^{} \vec \taub \cdot \vec n\, dS\\
\label{ang-vel}
I_p \td{\vec \omegab_p}{t} &= \rho_f \oint_{\partial \mathcal{V}_p}^{} \vec r \times \vec \taub \cdot \vec n\, dS
\end{align}
where $V_p = 4\upi a^3/3$ and $I_p=2 \rho_p V_p a^2/5$ are the particle volume and moment of inertia; $\vec \taub = -p \vec I + 2\mu \vec E$ 
is the fluid stress, with $\vec E = \left(\grad \vec u_f + \grad \vec u_f^T \right)/2$ the deformation tensor; $\vec r$ is the distance vector 
from the center of the sphere while $\bf{n}$ is the unity vector normal to the particle surface $\partial \mathcal{V}_p$. Dirichlet boundary 
conditions for the fluid phase are enforced on the particle surfaces as $\vec u_f|_{\partial \mathcal{V}_p} = \vec u_p + \vec \omegab_p \times \vec r$.

In the numerical code, an immersed boundary method is used to couple the fluid and solid phases. The boundary
condition at the moving particle surface (i.e. $\vec u_f|_{\partial \mathcal{V}_p} = \vec u_p + \vec \omegab_p \times \vec r$) is
modeled by adding a force field on the right-hand side of the Navier-Stokes equations. The fluid phase is therefore evolved in the whole 
computational domain using a second order finite difference scheme on a staggered mesh while the time integration is performed by a third 
order Runge-Kutta scheme combined with a pressure-correction method at each sub-step. The same integration scheme is also used for the Lagrangian 
evolution of eqs.~(\ref{lin-vel}) and (\ref{ang-vel}). Each particle surface is described by uniformly distributed $N_L$ Lagrangian points. 
The force exchanged by the fluid on the particles is imposed on each $l-th$ Lagrangian point and is related to the Eulerian force field 
$\vec f$ by the expression $\vec f(\vec x) = \sum_{l=1}^{N_L} \vec F_l \delta_d(\vec x - \vec X_l) \Delta V_l$. In the latter $\Delta V_l$ 
represents the volume of the cell containing the $l-th$ Lagrangian point while $\delta_d$ is the Dirac delta. This force field is calculated 
through an iterative algorithm that ensures a second order global accuracy in space. In order to maintain accuracy, eqs.~(\ref{lin-vel}) and 
(\ref{ang-vel}) are rearranged in terms of the IBM force field,
\begin{align}
\label{lin-vel-ibm}
\rho_p V_p \td{\vec u_p}{t} &= -\rho_f \sum_{l=1}^{N_l} \vec F_l \Delta V_l + \rho_f \td{}{t} \int_{\mathcal{V}_p}^{} \vec u_f\, dV \\
\label{ang-vel-ibm}
I_p \td{\vec \omegab_p}{t} &= -\rho_f \sum_{l=1}^{N_l} \vec r_l \times \vec F_l \Delta V_l + \rho_f \td{}{t} \int_{\mathcal{V}_p}^{} \vec r \times \vec u_f\, dV 
\end{align}
where $\vec r_l$ is the distance from the center of a particle while the second terms on the right-hand sides are corrections to account for 
the inertia of the fictitious fluid contained within the particle volume. Particle-particle interactions are also considered. When the gap 
distance between two particles is smaller than twice the mesh size, lubrication models based on Brenner's asymptotic solution \citep{brenner1961} 
are used to correctly reproduce the interaction between the particles. A soft-sphere collision model is used to account for collisions between particles 
and between particles and walls. An almost elastic rebound is ensured with a restitution coefficient set at $0.97$. These lubrication and collision 
forces are added to the right-hand side of eq.~(\ref{lin-vel-ibm}). For more details and validations of the numerical code, the reader is referred 
to previous publications \cite{breugem2012,lambert2013,picano2015,fornari2015}.

\subsection{Flow configuration}

We consider a turbulent channel flow between two infinite flat walls located at $y=0$ and $y=2h$, where $y$ is the wall-normal direction while 
$x$ and $z$ are the streamwise and spanwise directions. The domain has size $L_x=6h$, $L_y=2h$ and $L_z=3h$ and periodic 
boundary conditions are imposed in the streamwise and spanwise directions. A fixed value of the bulk velocity $U_0$ is achieved by imposing 
a mean pressure gradient in the streamwise direction. The imposed bulk Reynolds number is equal to $Re_b=U_02h/\nu=5600$ (where $\nu$ represents 
the kinematic viscosity of the fluid) and corresponds to a Reynolds number based on the friction velocity $Re_{\tau}=U_*h/\nu=180$ for the 
single phase case. The friction velocity is defined as $U_*=\sqrt{\tau_w/\rho_f}$, where $\tau_w$ is the stress at the wall. A cubic staggered 
mesh of $864\times288\times432$ grid points is used to discretize the domain. All results will be reported either in 
non-dimensional outer units (scaled by $U_0$ and $h$) or in inner units (with the superscript '+', scaled by $U_*$ and $\delta_*=\nu/U_*$).

\begin{figure}
\centering
\includegraphics[scale=0.35,angle=270]{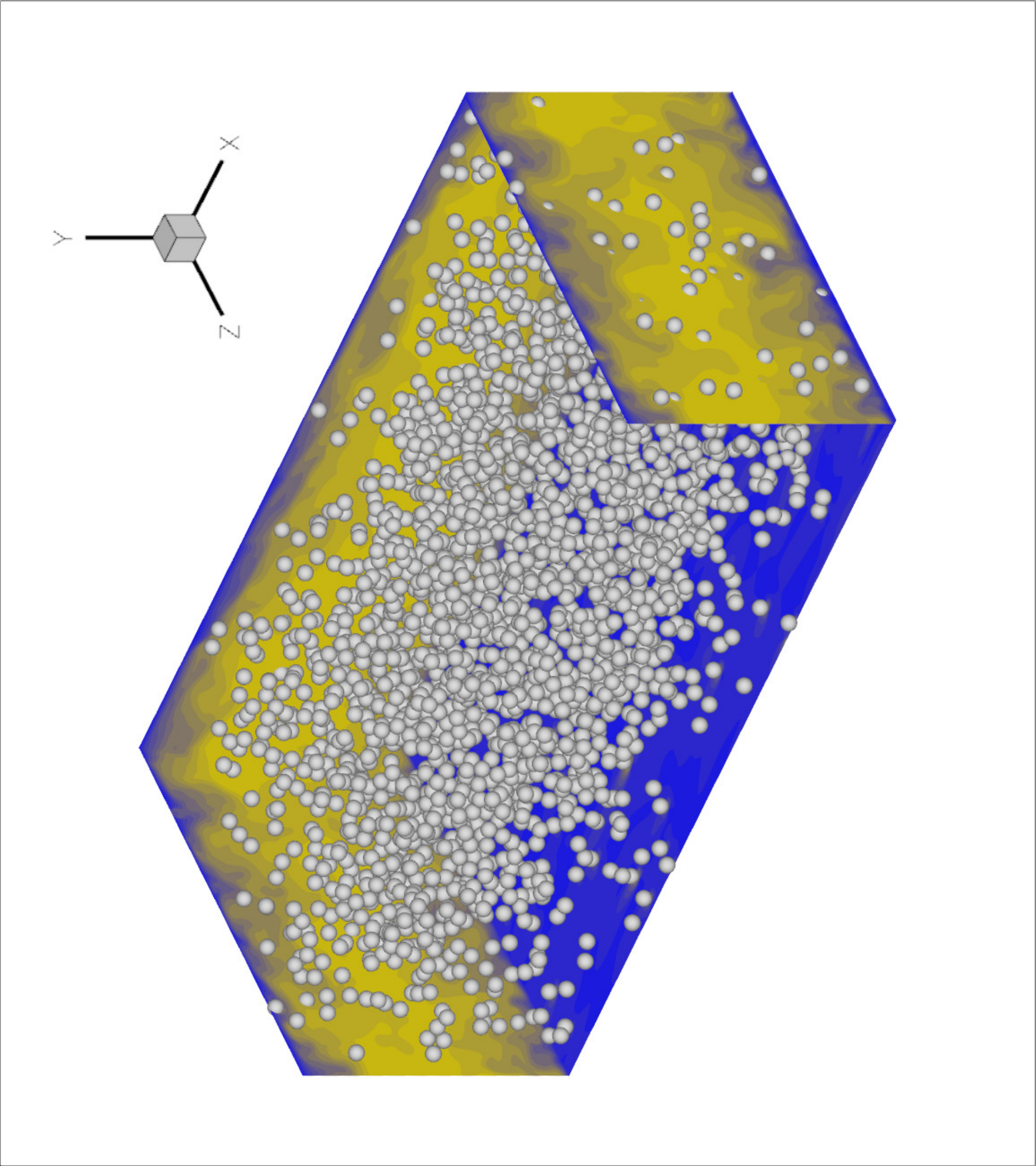}
\caption{Instantaneous snapshot of the streamwise velocity on different orthogonal planes together with the corresponding particle position, for $R=10$. \label{fig:config}}
\label{fig:snap}
\end{figure}

The solid phase consists of non-Brownian rigid spheres with a radius to channel half-width ratio fixed to $a/h=1/18$. For a 
volume fraction $\phi=5\%$, this radius corresponds to about 10 plus units. In figure~\ref{fig:snap} we display the instantaneous streamwise velocity on four 
orthogonal planes together with the finite-size particles dispersed in the domain. Each particle is discretized with $N_l=746$ Lagrangian control points 
while their radii are  $8$ Eulerian grid points long. Using an Eulerian mesh consisting of $8$ grid points per particle radius ($\Delta x = 1/16$) is a good compromise in terms 
of computational cost and accuracy. We have performed a simulation with a finer mesh ($12$ points per particle radius, $\Delta x = 1/24$), $R=1$ and $\phi=5\%$. We 
find indeed that the friction Reynolds number $Re_{\tau}$ changes by $1\%$, and the velocity fluctuations change locally at most by $4\%$.

At first, we will compare results obtained at different density ratios $R$ and constant mass fraction $\chi$ with those at constant volume fraction $\phi$. The 
mass fraction is defined as $\chi=\phi R$ and is chosen to be $0.2$: four simulations are performed with $R=1, 4, 10, 100$ and 
$\phi=20\%, 5\%, 2\%, 0.2\%$ (which correspond to $10000,2500,1000$ and $100$ particles). At constant $\phi=5\%$ instead, 
we examine four cases with $R=1, 2, 4$ and $10$. The reference unladen case ($\phi=0\%$) is also presented in the different figures. The case with $\phi=5\%$ and $R=1000$ 
will be discussed later. The full set of simulations  is summarized in table~\ref{tab:sim}.

The simulations start from the laminar Poiseuille flow for the fluid phase since we observe that the transition naturally 
occurs at the present moderately high Reynolds number, due to the noise added by the particles. 
Particles are initially positioned randomly 
with velocity equal to the local fluid velocity. Statistics are collected after the initial transient phase.

\begin{table}
  \begin{center}
\def~{\hphantom{0}}
  \begin{tabular}{cccc}
     $\phi ($\%$)$  & $N_p$    & $\chi$   &  $R$    \\[3pt]
     $0$            & $0   $   & $0$      &  $--$   \\
     $0.2$          & $101 $   & $0.2$    &  $100$  \\
     $2$            & $1000$   & $0.2$    &  $10$   \\
     $5$            & $2500$   & $0.05$   &  $1$    \\
     $5$            & $2500$   & $0.1$    &  $2$    \\
     $5$            & $2500$   & $0.2$    &  $4$    \\
     $5$            & $2500$   & $0.5$    &  $10$   \\
     $5$            & $2500$   & $50$    &  $1000$   \\
     $20$           & $10000$  & $0.2$    &  $1$    \\
  \end{tabular}
  \caption{Summary of the simulations performed ($N_p$ is the total number of particles).}
  \label{tab:sim}
  \end{center}
\end{table}

\begin{figure}
\centering
\includegraphics[width=.50\textwidth]{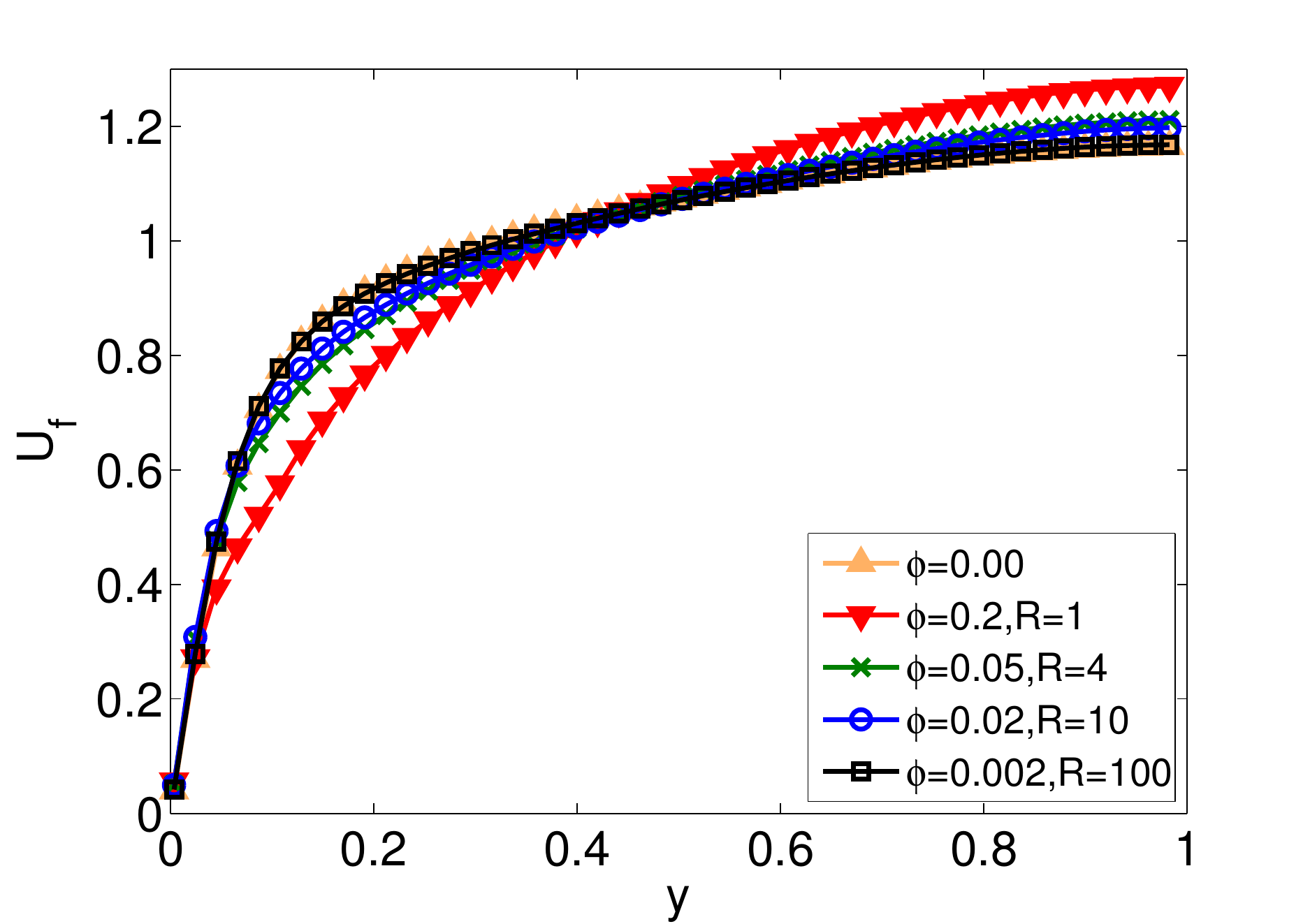}~~~~
\put(-205,110){{\large a)}}
{\includegraphics[width=.50\textwidth]{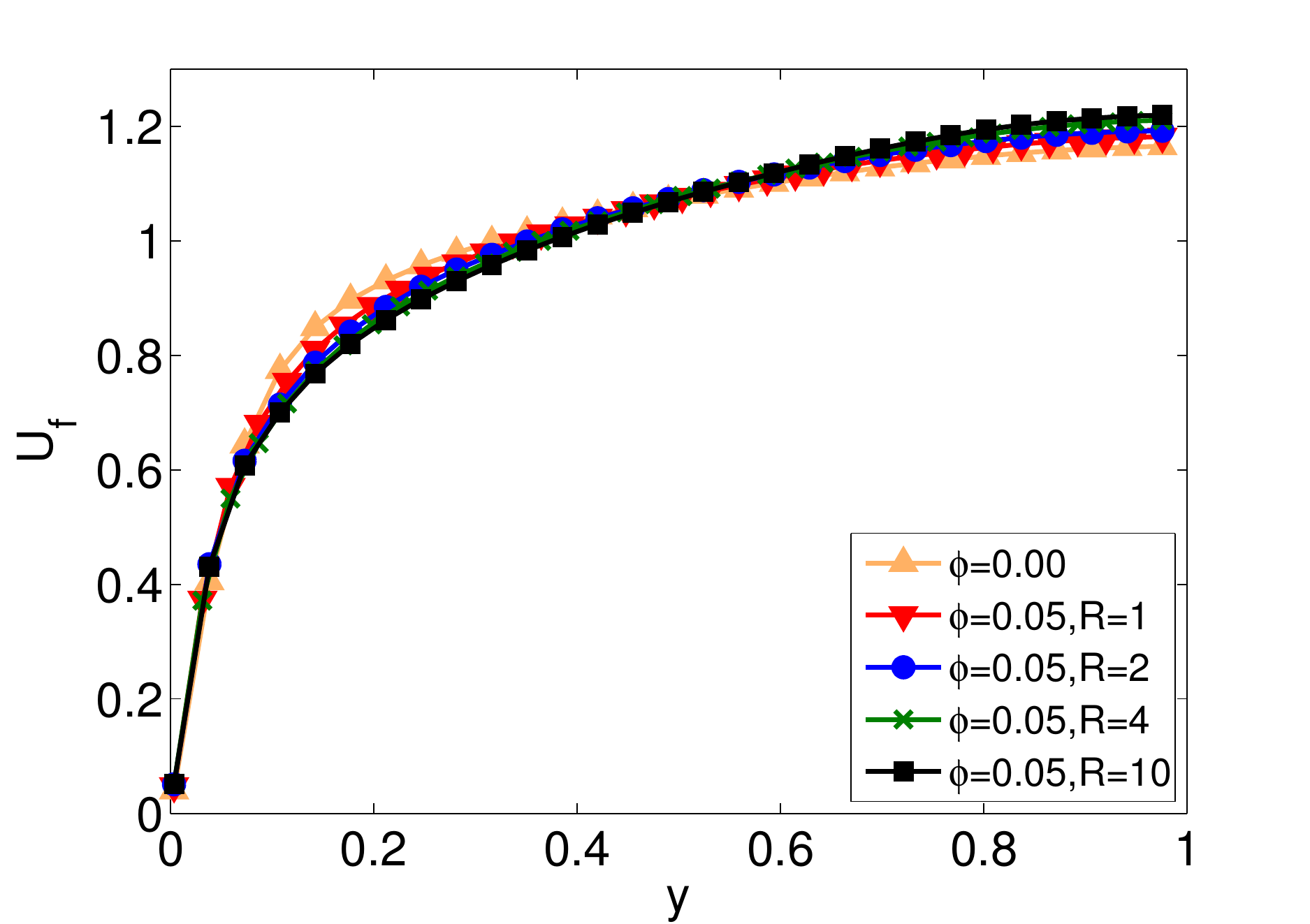}
\put(-205,110){{\large b)}}}
\includegraphics[width=.50\textwidth]{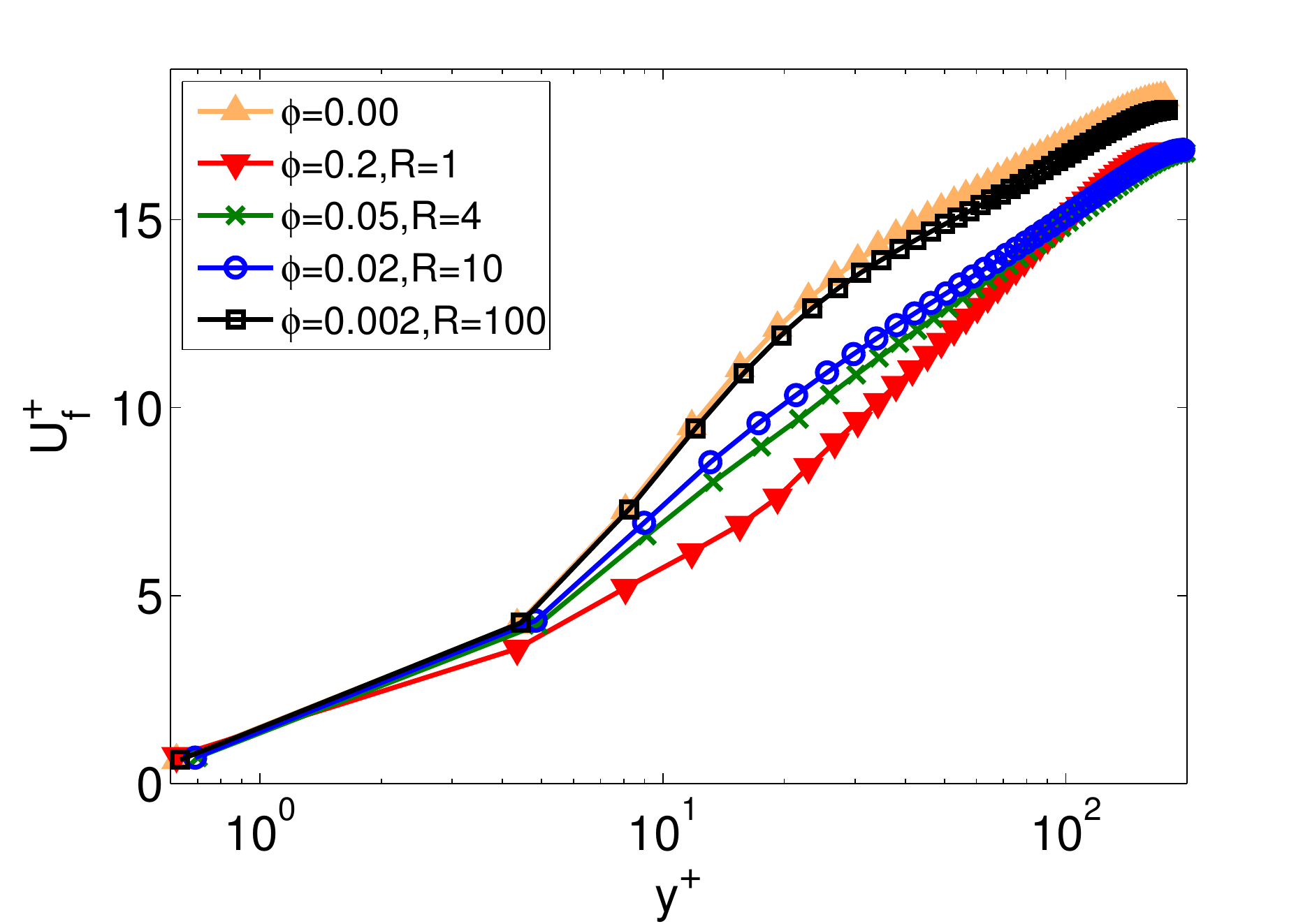}~~~~
\put(-205,110){{\large c)}}
{\includegraphics[width=.50\textwidth]{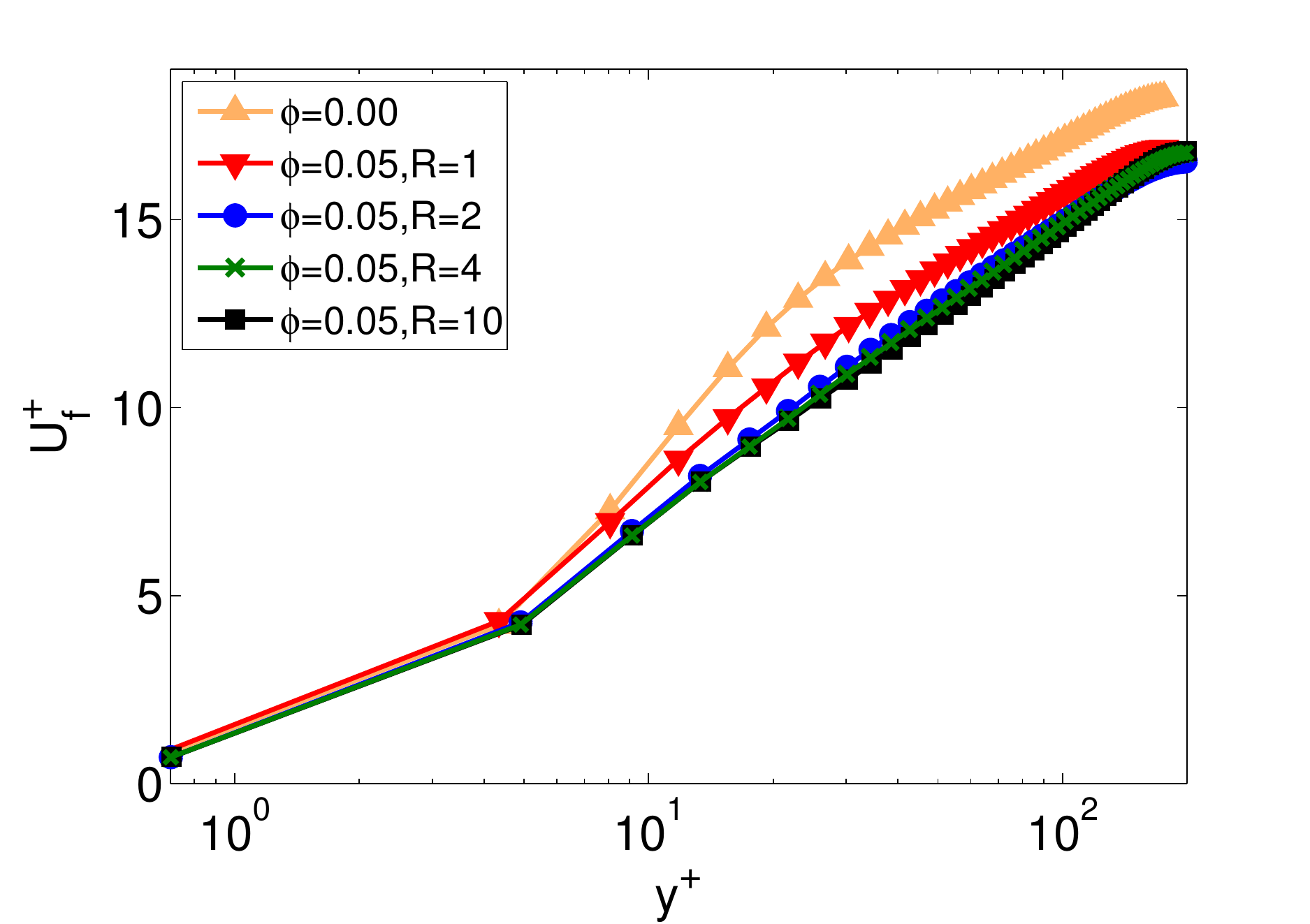}
\put(-205,110){{\large d)}}}
\caption{Mean fluid streamwise velocity profiles for constant $\chi=0.2$;  (a) data scaled  in outer units and  (c) inner units. The corresponding cases at constant $\phi=0.05$ are displayed in (b) and (d).}
\label{fig:Uf}
\end{figure}

\section{Results}

\subsection{Analysis of Mass and Volume Fraction Effects}

We show the mean fluid velocity profiles in outer and inner units ($U_f^+=U_f/U_*$ and $y^+=y/\delta_*$) in figure~\ref{fig:Uf}. 
The statistics conditioned to the fluid phase have been calculated neglecting the points occupied by the solid phase in each field 
(phase-ensemble average). 
We notice in~\ref{fig:Uf}(a) and (c) that  the velocity profile tends towards 
that for the single fluid phase as the volume fraction is reduced even if the mass fraction $\chi$ is constant. 
Conversely, when the volume fraction is kept constant at $5\%$ (panels b and 
d) the differences observed when increasing the density ratios are small; in particular smaller velocities near the wall and 
larger velocities in the centre of the channel for larger $\chi$. The decrease of the profiles in inner units, observed when increasing $\phi$ and 
less so increasing $R$ at fixed $\phi$, indicates also an overall 
drag increase. Indeed for $y^+ > 40-50$ the mean profile follows the log-law \cite{pope2000}:
\begin{equation}
U^+ = \frac{1}{k} log\left(y^+\right) + B 
\label{loglaw}
\end{equation}
where $k$ and $B$ are the von K\'{a}rm\'{a}n constant and an additive coefficient. As $R$ increases, $k$ is found to decrease from $0.36$ to 
$0.29$ while $B$ is reduced from $2.7$ to $-1.3$ (see figure~\ref{fig:Uf}d). Usually a decrease in $k$ denotes drag reduction while a smaller or negative 
$B$ leads to an increase in drag \cite{virk1975}. In the cases studied this combined effect leads to a small increase of the overall drag since the 
friction Reynolds number $Re_{\tau}$ grows from $195$ to $203$. The reduction in the additive coefficient $B$ is believed to be caused by the 
intense particle-fluid interactions occurring near the wall \cite{picano2015}, which are augmented by the increased inertia of the solid 
phase at higher $R$.

\begin{table}
  \begin{center}
\def~{\hphantom{0}}
  \begin{tabular}{cccccc}
      $\chi$ & $\phi ($\%$)$  & $R$   &   $k$ & $B$ & $Re_{\tau}$ \\[3pt]
      $0$    & $0$            & $--$  & $0.40$ & $5.5$ & $180$ \\
      $0.2$  & $0.2$          & $100$ & $0.38$ & $4.7$ & $183$ \\
      $0.2$  & $2$            & $10$  & $0.33$ & $1.1$ & $198$ \\
      $0.05$ & $5$            & $1$   & $0.36$ & $2.7$ & $195$ \\
      $0.1$  & $5$            & $2$   & $0.33$ & $1.0$ & $201$ \\
      $0.2$  & $5$            & $4$   & $0.30$ & $-0.3$& $202$ \\
      $0.5$  & $5$            & $10$  & $0.29$ & $-1.3$& $203$ \\
      $0.2$  & $20$           & $1$   & $0.22$ & $-6.3$ & $216$ \\
  \end{tabular}
  \caption{Summary of the values of the von K\'{a}rm\'{a}n constant $k$, the additive coefficient $B$ and the friction Reynolds number $Re_{\tau}$ 
obtained for the cases studied. The reference case with no dispersed phase is also reported. Here $k$ and $B$ have been fitted in the range $y^+ \in [50,150]$.}
  \label{tab:kBRe}
  \end{center}
\end{table}
\begin{figure}
\centering
\includegraphics[width=.50\textwidth]{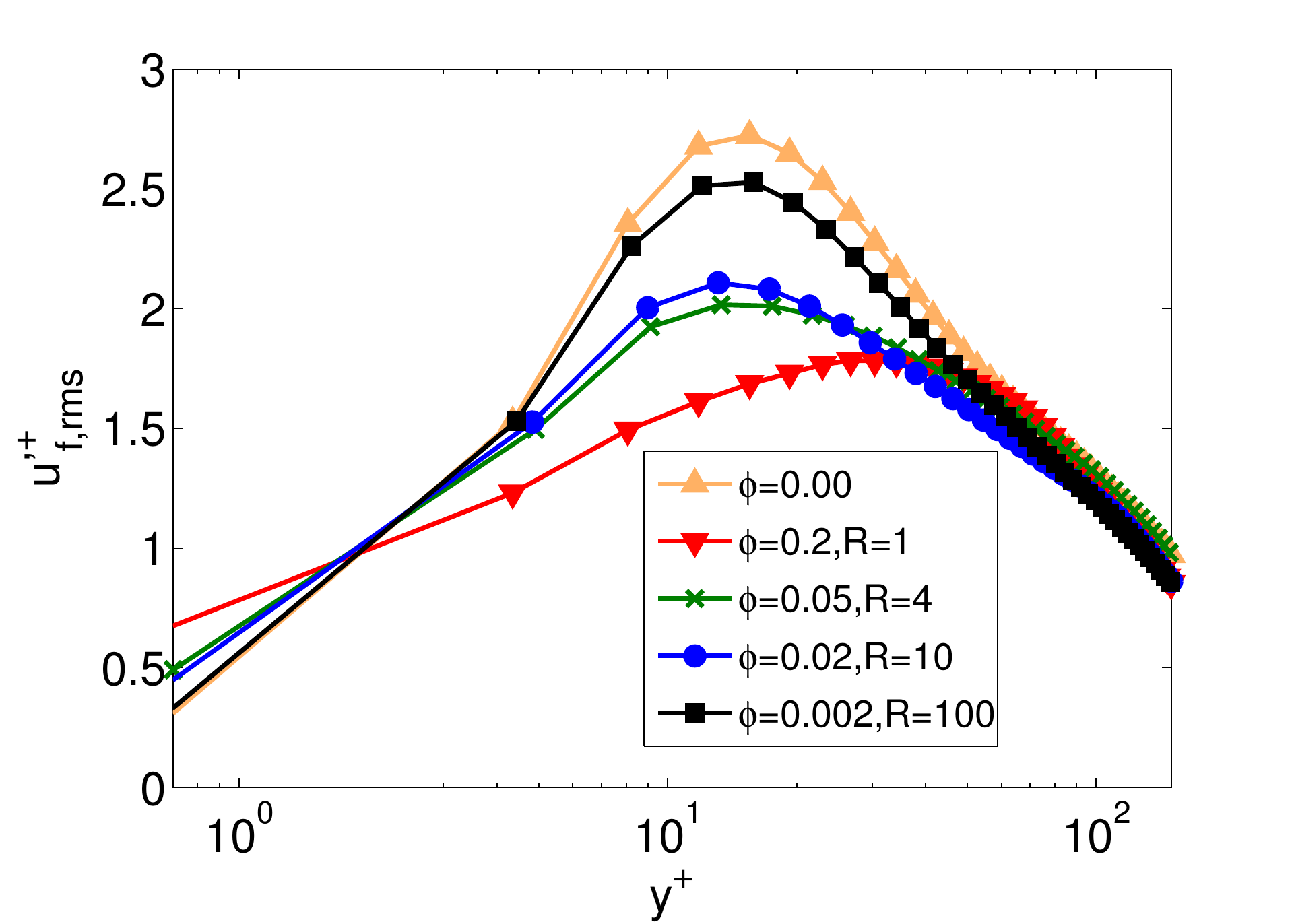}~~~~
\put(-205,110){{\large a)}}
{\includegraphics[width=.50\textwidth]{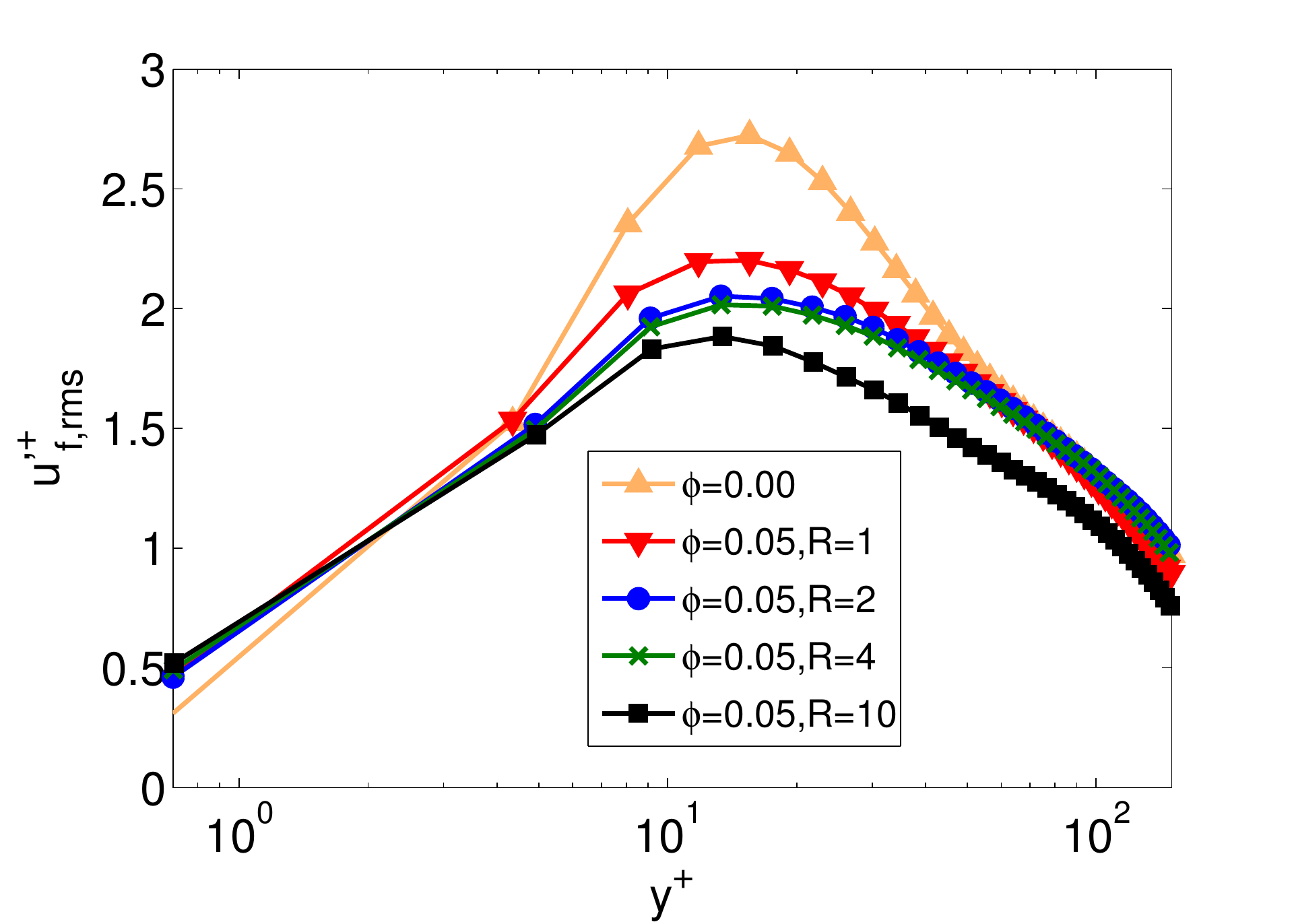}
\put(-205,110){{\large b)}}}
\includegraphics[width=.50\textwidth]{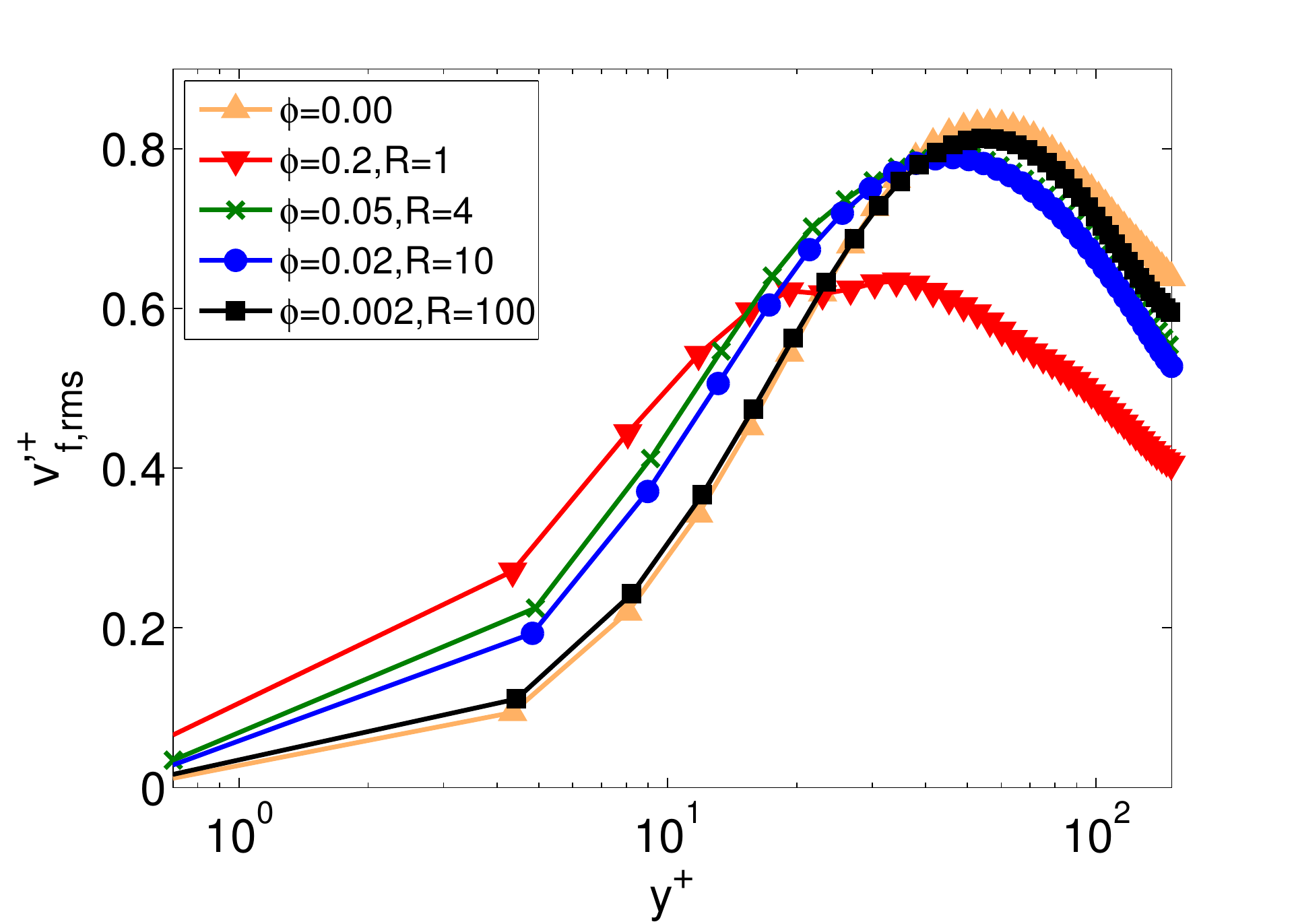}~~~~
\put(-205,110){{\large c)}}
{\includegraphics[width=.50\textwidth]{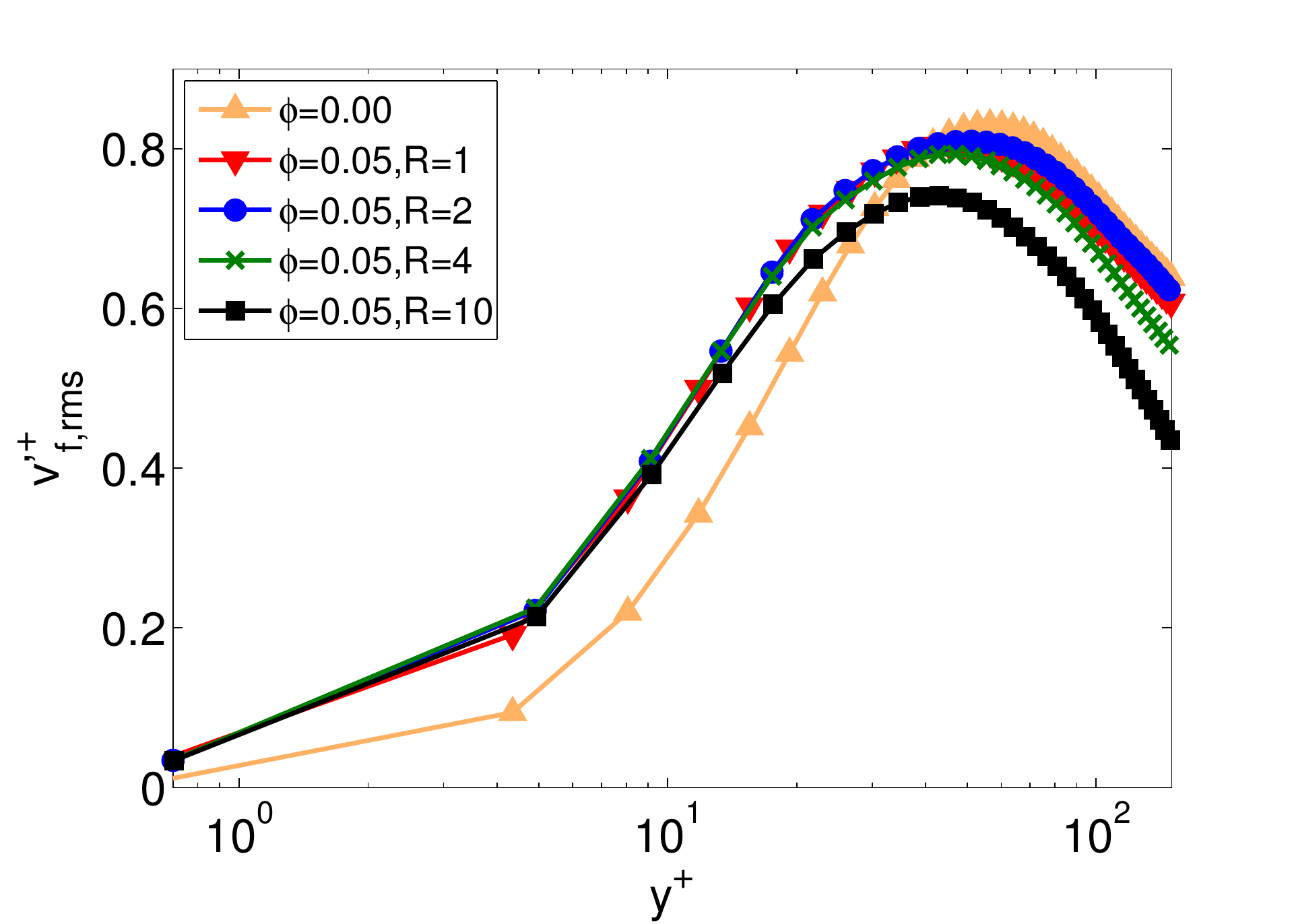}
\put(-205,110){{\large d)}}}
\includegraphics[width=.50\textwidth]{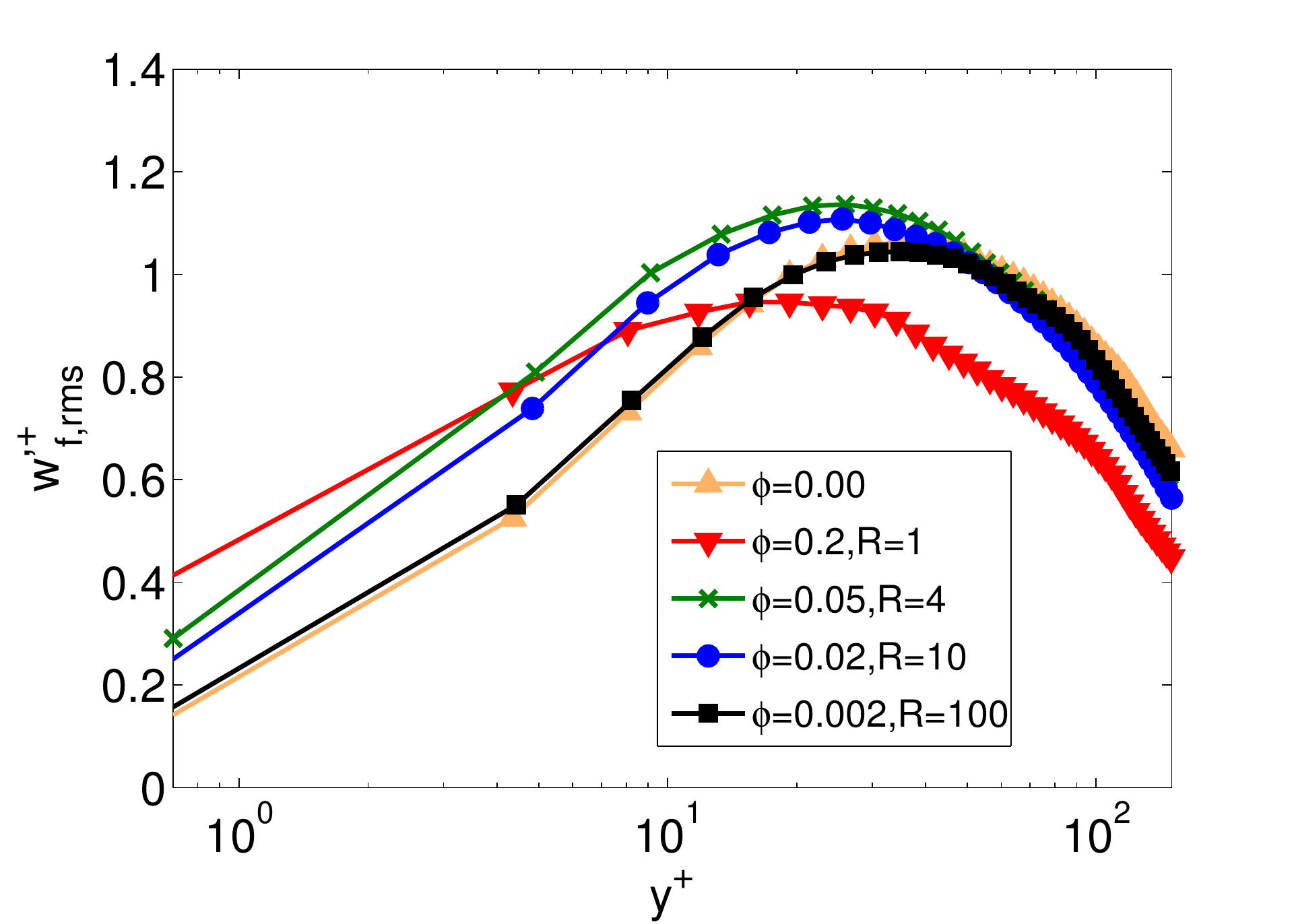}~~~~
\put(-205,110){{\large e)}}
{\includegraphics[width=.50\textwidth]{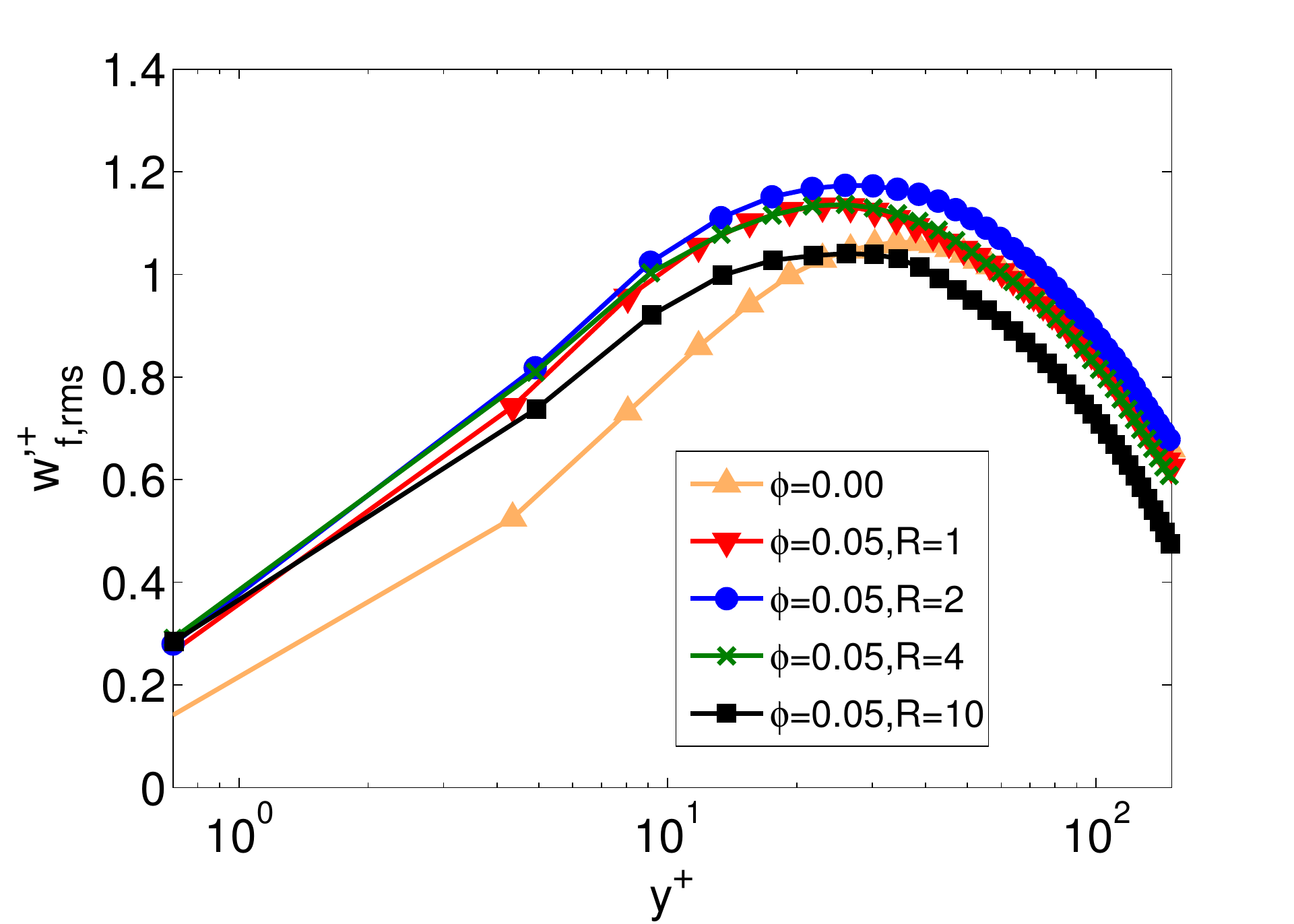}
\put(-205,110){{\large f)}}}
\caption{Intensity of the fluctuation velocity components for the fluid phase in inner units. (a),(c),(e) simulations at constant mass fraction $\chi=0.2$. (b),(d),(f)  Data at constant volume fraction $\phi=0.05$.}
\label{fig:Urf}
\end{figure}

We report in table~\ref{tab:kBRe} the values of $k, B$ and $Re_{\tau}$ obtained for all the cases studied. For the case with $\phi=0.2\%$ and 
$R=100$ (yet $\chi = 0.2$) we almost recover the single phase log-law with $k=0.38$ and $B=4.7$ (for the single fluid $k=0.4$ and $B=5.5$) and the 
increase in friction Reynolds number $Re_{\tau}$ is limited (from $180$ to $183$), which can be explained by the small number of particles in the flow. 
As shown above, the cases at the same mass fraction ($\chi = 0.2$) and different density ratios reveal most 
significant variations, explained by the changes in volume fraction $\phi$ (excluded volume effect).

At high $\phi$, the bulk flow seems to be relaminarized, see figure~\ref{fig:Uf}(a). However, although the Reynolds stresses are reduced, 
the particle presence induces fluctuations and enhances the overall friction via particle-induced stresses \cite{lashgari2014,picano2015,lash2015}.

The root-mean-square (r.m.s.) of the fluid velocity fluctuations are reported in inner units in figure~\ref{fig:Urf}. Panels (a),(c) and (e) 
show the cases at constant $\chi$ while the cases at constant $\phi$ are reported in panels (b),(d) and (f). As for the mean flow, the major 
changes in fluid velocity fluctuations are associated to an increase in volume fraction $\phi$. As $\phi$ is increased from $0.2\%$ to $5\%$ (constant $\chi$), the 
wall-normal $v_{f,rms}'^{+}$ and spanwise components $w_{f,rms}'^{+}$ increase especially in the proximity of the wall, i.e. in the 
viscous sublayer. We observe also an important reduction of the streamwise fluctuation intensity around $y^+=10$ at higher $\phi$. 
As we will show later, a layer of particles is formed close to the walls and the fluid between these particles and the walls is therefore squeezed. 
This results in a reduction of the streamwise fluid velocity fluctuations and an increase of the fluctuations in the other directions.
The neutrally buoyant case at higher volume fractions ($\phi=20\%$, $R=1$) exhibits higher fluctuations close to the walls that drop well below the values 
found for the smaller volume fractions $\phi$ as $y^+$ is further increased. Only the streamwise component $u_{f,rms}'^{+}$ approaches the values 
obtained at smaller $\phi$ when $y^+ > 80$.

The fluid velocity fluctuation profiles do not show a significant dependence on the density ratio $R$. However, one can notice that increasing the 
density ratio to $R=10$ leads to a reduction of the fluctuation intensities in all directions (when $y^+ > 5$), similarly to what 
observed at $R=1$ and increasing $\phi$ (see previous discussion or the work by Picano and collaborators \cite{picano2015} for a 
more complete discussion). 
Important differences are found for $y^+ < 5$ (i.e. 
very close to the wall) where the velocity fluctuations increase when increasing the volume fraction while they remain approximately 
constant when varying $R$.

\begin{figure}
\centering
\includegraphics[width=.50\textwidth]{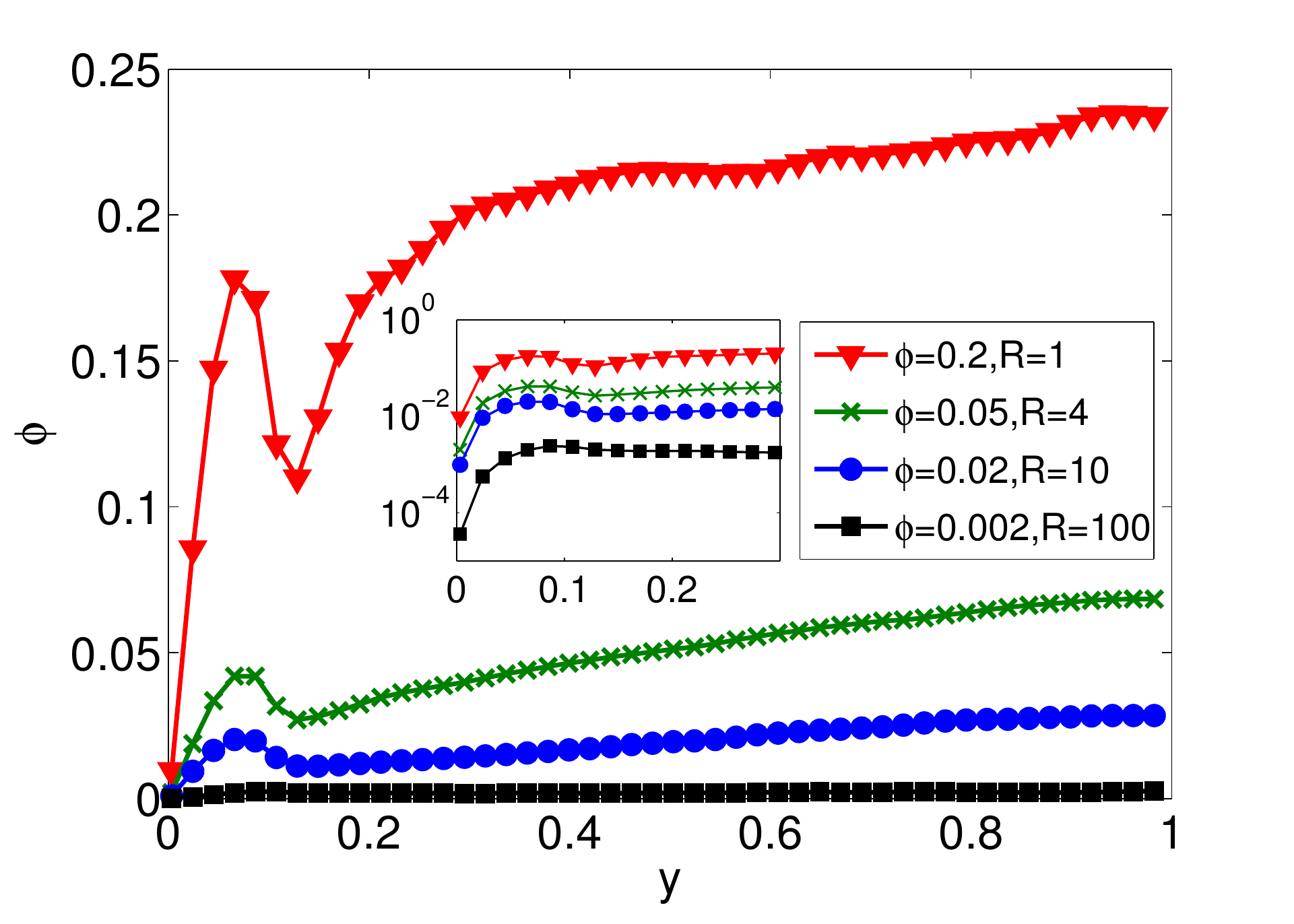}~~~~
\put(-205,110){{\large a)}}
{\includegraphics[width=.50\textwidth]{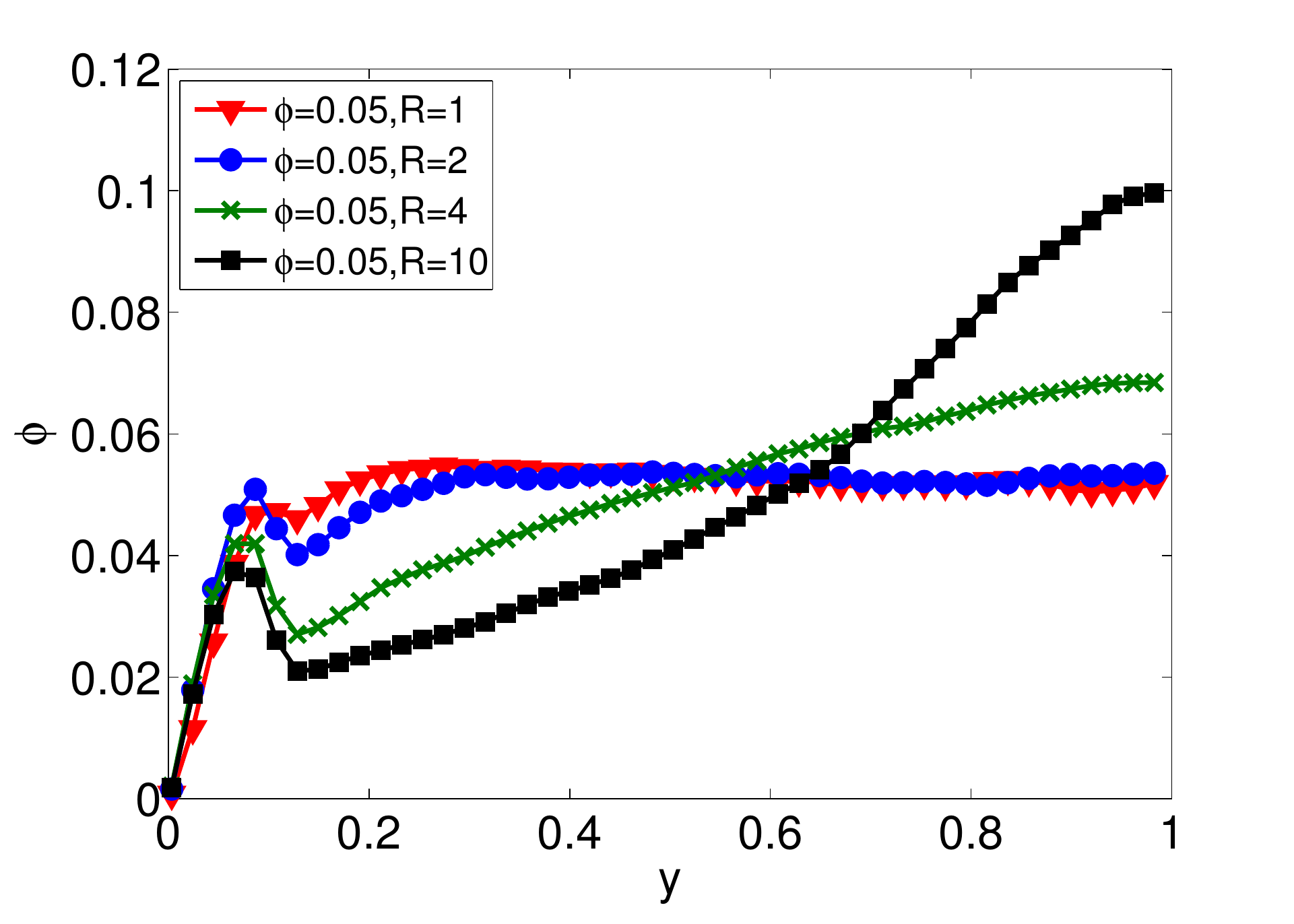}
\put(-205,110){{\large b)}}}
\caption{Mean local volume fraction $\phi$ versus wall-normal coordinate $y$. (a) constant $\chi=0.2$ and (b)  constant $\phi=0.05$.}
\label{fig:phi1}
\end{figure}

As mean velocity profiles are affected mostly by variations in the solid volume fraction $\phi$, the 
explanation for the change in fluid velocity fluctuations must be searched in the context of fluid-solid interactions and of particle distribution. 
We therefore report in figure~\ref{fig:phi1} the local solid volume fraction along the wall-normal direction $\phi(y)$. The phase-ensemble 
averages for the solid phase have been obtained considering the Eulerian grid points contained within the volume of each particle at each time step.
It is evident that for $R > 1$ a layer of particles forms close to the walls as soon as the volume fraction $\phi$ is above $0.2\%$. 

As shown in figure~\ref{fig:phi1}(b) for a constant volume fraction ($\phi=5\%$), 
as the density ratio increases 
more particles tend to migrate toward the centerline while 
the layer close to the wall is preserved. 
 The peak of $\phi(y)$ close to the wall is 
slightly reduced and less particles occupy the volume between $y \sim 0.1$ and $0.6$. We therefore observe a shear-induced particle migration 
from regions of high to low shear rates, 
an effect more pronounced as the density ratio $R$ increases. 
The local volume fraction 
increases drastically at the centerline ($y = 1$): 
the local volume fraction at the centerline $\phi(y = 1)$ is approximately twice that 
 found at $y \sim 0.1$ (i.e. close to the wall where the first layer of particles form) when $R=4$. The difference is even higher when $R=10$-- $\phi(y = 1) \simeq 5\phi(y = 0.1)$. 
This shear-induced migration becomes more intense as the density ratio $R$ increases although, as we will see later, the picture totally 
changes at very high $R$ ($\sim 1000$).

\begin{figure}
\centering
\includegraphics[width=.50\textwidth]{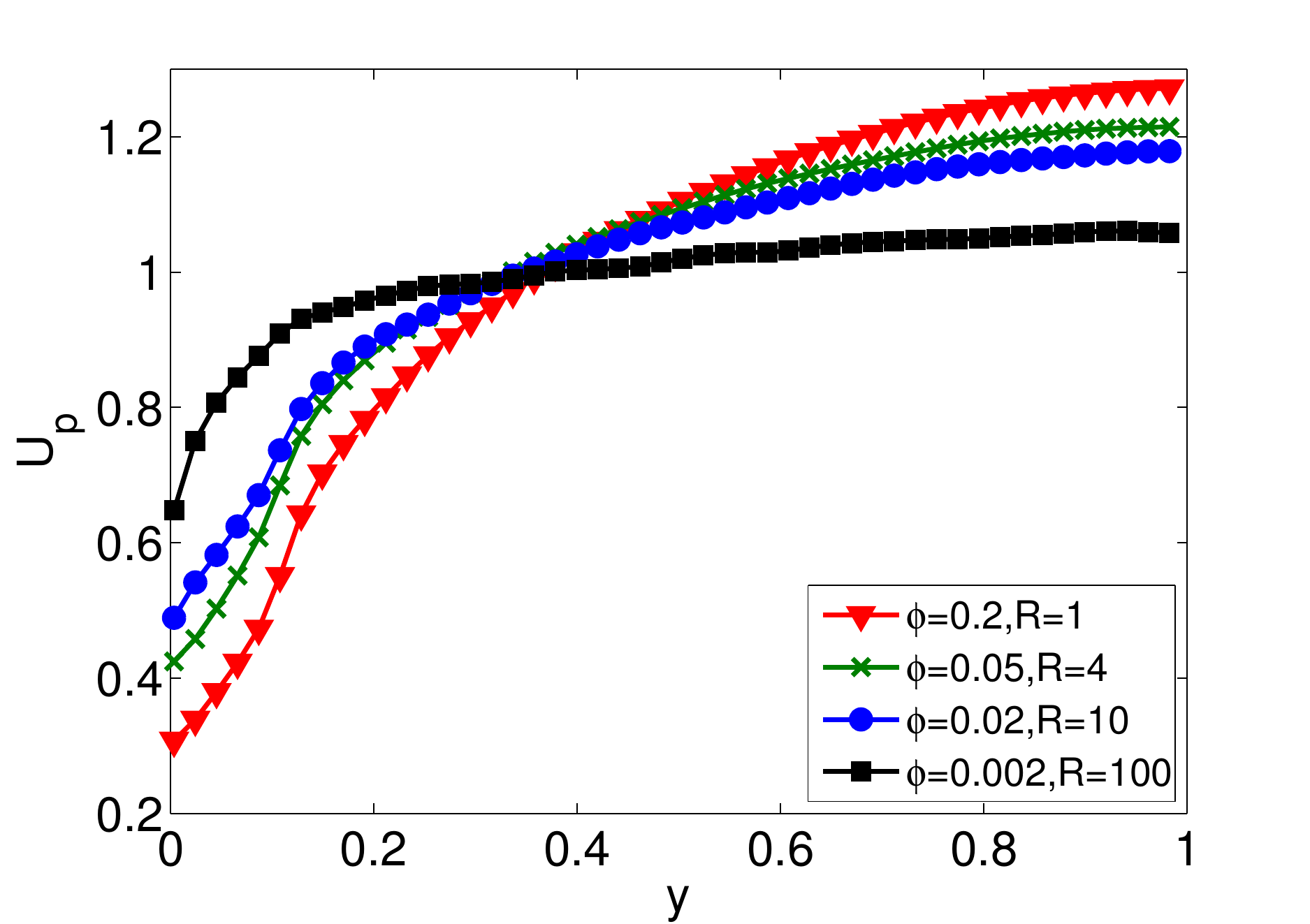}~~~~
\put(-205,110){{\large a)}}
{\includegraphics[width=.50\textwidth]{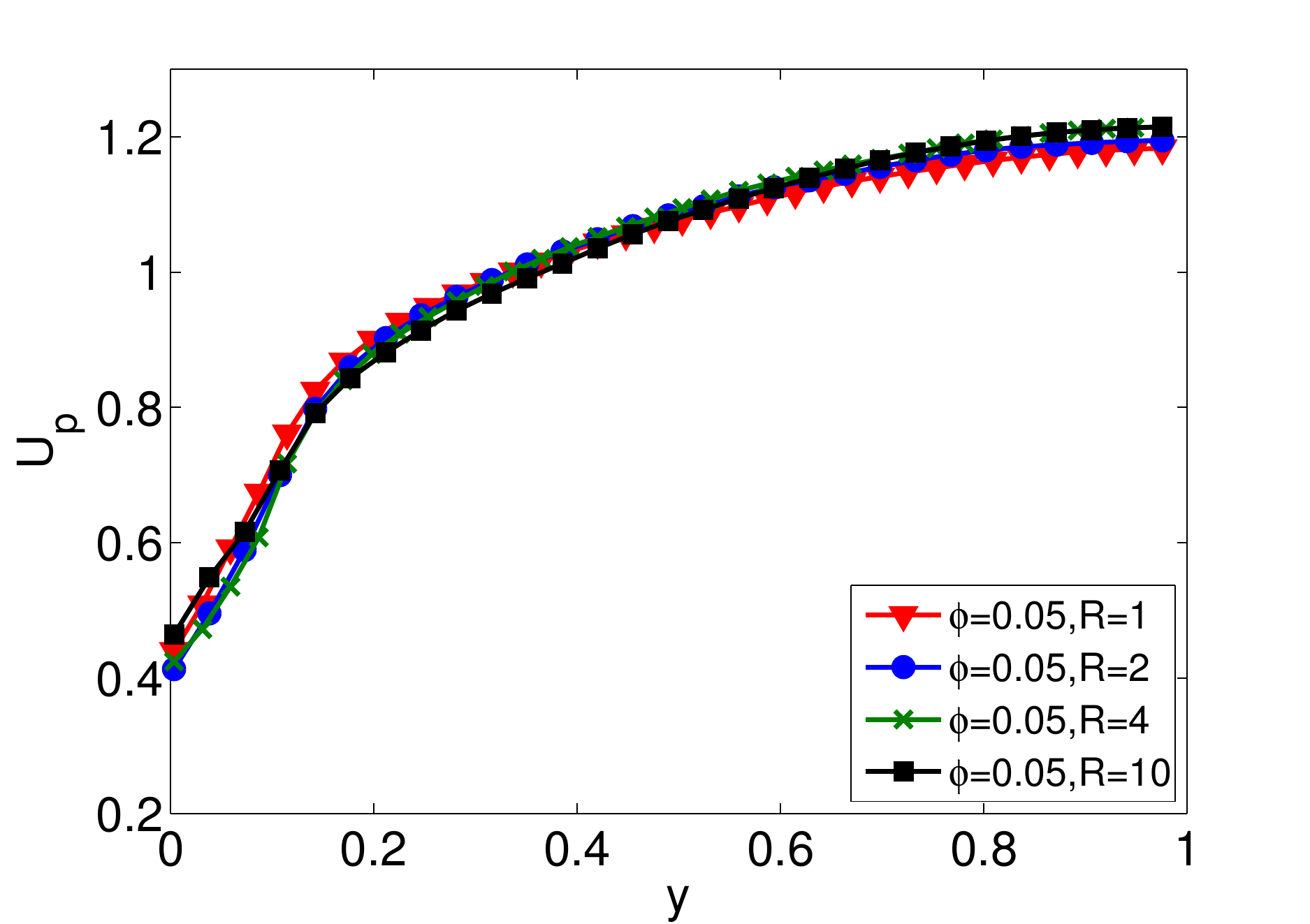}
\put(-205,110){{\large b)}}}
\caption{Mean particle streamwise velocity profiles for (a) constant $\chi=0.2$ and (b) constant $\phi=0.05$.}
\label{fig:Up1}
\end{figure}

We report the mean particle streamwise velocity $U_p$ in figure~\ref{fig:Up1}. The results for constant $\chi$ are shown in panel (a) where we notice that 
the mean particle streamwise velocity profiles are similar for $\phi=2\%$ and $5\%$ when $R=10$ and $4$.
For $\phi=0.2\%$ and $R=100$ instead, the 
mean particle streamwise velocity profile changes drastically showing higher velocities close to the wall and smaller velocities in the rest of the 
channel ($y \ge 0.3$). Comparing with the other cases we find a $13\%$ reduction of $U_p$ at the centerline. Generally we find that as the volume 
fraction $\phi$ increases, the mean particle streamwise velocity decreases closer to the walls while exhibiting higher values at the centerline.

When the volume fraction $\phi$ is fixed (fig.~\ref{fig:Up1}b), $U_p$ is only slightly altered by an increase in density ratio $R$. However, 
at the highest density ratio ($R=10$) particles move 
faster in proximity of the walls and around the centerline while $U_p$ is reduced between these two regions.
The particles that lie in this region have 
a streamwise velocity directly linked to that of the fluid, while particles are accelerated in proximity of the wall and around the centerline where collision are more frequent.
The mean particle velocity is finite close to the walls, since particles can have a relative tangential motion. The phase-ensemble averages of particle 
velocities are computed considering the velocities of the Eulerian grid points contained within the volume of each particle, with $\vec u(\vec X,t)=\vec u_p^q(\vec X_p^q,t) + 
\vec \omegab_p^q(t) \times \left(\vec X - \vec X_p^q(t) \right) $ and $\vec X \in [\vec X_p^q(t) \pm a]$ (where $\vec X_p^q$ and $\vec u_p^q$ are the position of the $q-th$ 
particle centroid and its velocity).

\begin{figure}
\centering
\includegraphics[width=.50\textwidth,angle=270]{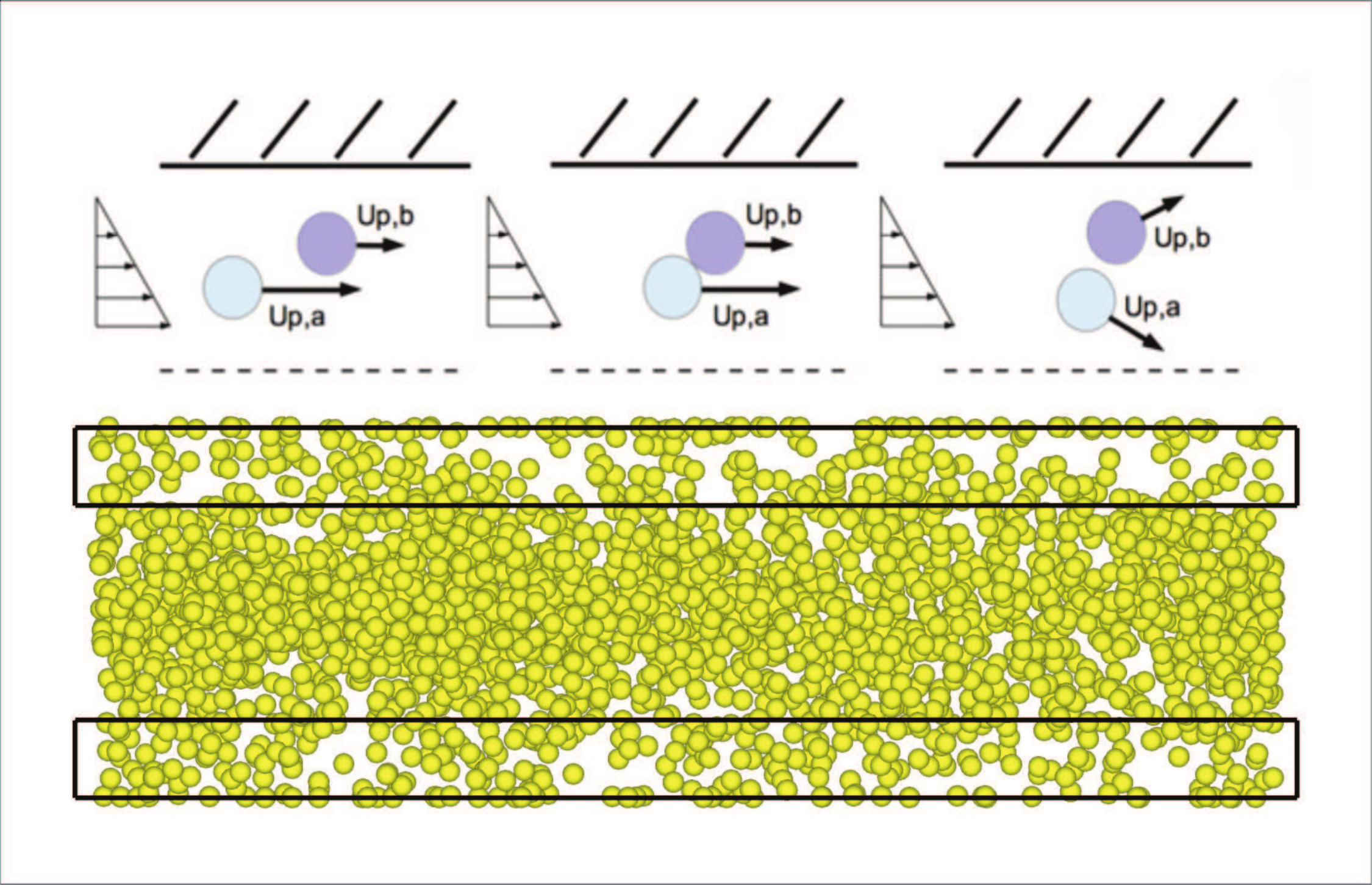}
\caption{Instantaneous particle positions in the $x-y$ plane from the simulation with $R=10$. On top, a sketch  
explaining the observed shear-induced migration is also presented.}
\label{fig:phi2}
\end{figure}

In figure~\ref{fig:phi2} we show the instantaneous particle positions from the simulation with $R=10$ projected in the streamwise-wall-normal 
($x-y$) plane. The interaction between two approaching particles slightly shifted in the wall-normal direction and in the proximity of the wall 
is also sketched to explain shear-induced inertial migration. 
In this high shear rate region, the particle denoted by $a$, with velocity $U_{p,a}$, approaches  particle $b$, moving in the same direction
 with velocity $U_{p,b}$. Since the latter is closer to the wall, its streamwise velocity $U_{p,b}$ is smaller (on average) than that of particle 
$a$, so a collision takes place. The scenario following this collision depends on the inertia of the 
fluid and solid phases, and thereby on the density ratio $R$.

If particles $a$ and $b$ are neutrally buoyant $R=1$, their dynamics is mainly determined by the carrier fluid flow. After the collision, 
the two particles would tend to move radially apart and their motion becomes rapidly correlated to that of the fluid phase. As a result, they are on average transported 
 downstream by the flow. As the particle inertia increases (i.e.\ $R$ increases), the particle motion is less sensitive to the fluid flow and longer 
times are needed for the particle to adjust to the fluid velocity after the collision. Indeed, for  $R \ge 4$, the particle relaxation time 
is  longer than the fluid timescale. Therefore, particles migrate almost undisturbed in opposite wall-normal directions after a collision. 
Owing to the presence of the wall, we therefore observe a net migration towards the channel centre.
Being this an inertial effect, the particle migration is more evident as the solid to fluid density ratio $R$ increases. 
As we will discuss later, however, this effect disappears at very high density rations, $R$, when the particle mean velocity is almost uniform, and there is no  a mean shear.
On average, this inertial shear-induced migration leads to high peaks of the local solid volume fraction $\phi(y)$ at the centerline (see figure~\ref{fig:phi1}
b). The effect is so strong at $R=10$, that it is easy to identify intermittently depleted regions of particles close to the walls (as shown in the snapshot 
in figure~\ref{fig:phi2}).

A similar wall-normal particle migration has been observed for dense suspensions ($\phi=30\%$) of neutrally buoyant rigid spherical particles at bulk Reynolds 
numbers $Re_b$ ranging from $500$ to $5000$ \cite{lash2015}. In these cases, the profiles of local volume fraction, $\phi(y)$, do not vary significantly by 
increasing the bulk Reynolds number and the observed migration has been attributed to the imbalance of normal stresses in the wall-normal direction.
Although the resulting behavior is similar, the driving mechanisms are different.

In this section we have studied the dependence of the suspension properties on both the solid to fluid density ratio $R$ and the solid volume fraction 
$\phi$. We have shown that the mean and fluctuating velocity fields of both phases are predominantly influenced by variations in the volume fraction $\phi$ 
(i.e. excluded volume effects). The mean fields are only marginally altered by increasing the density ratio $R$. The main effect of increasing particle inertia 
is the shear-induced migration just discussed.

\begin{figure}
\centering
\includegraphics[width=.50\textwidth]{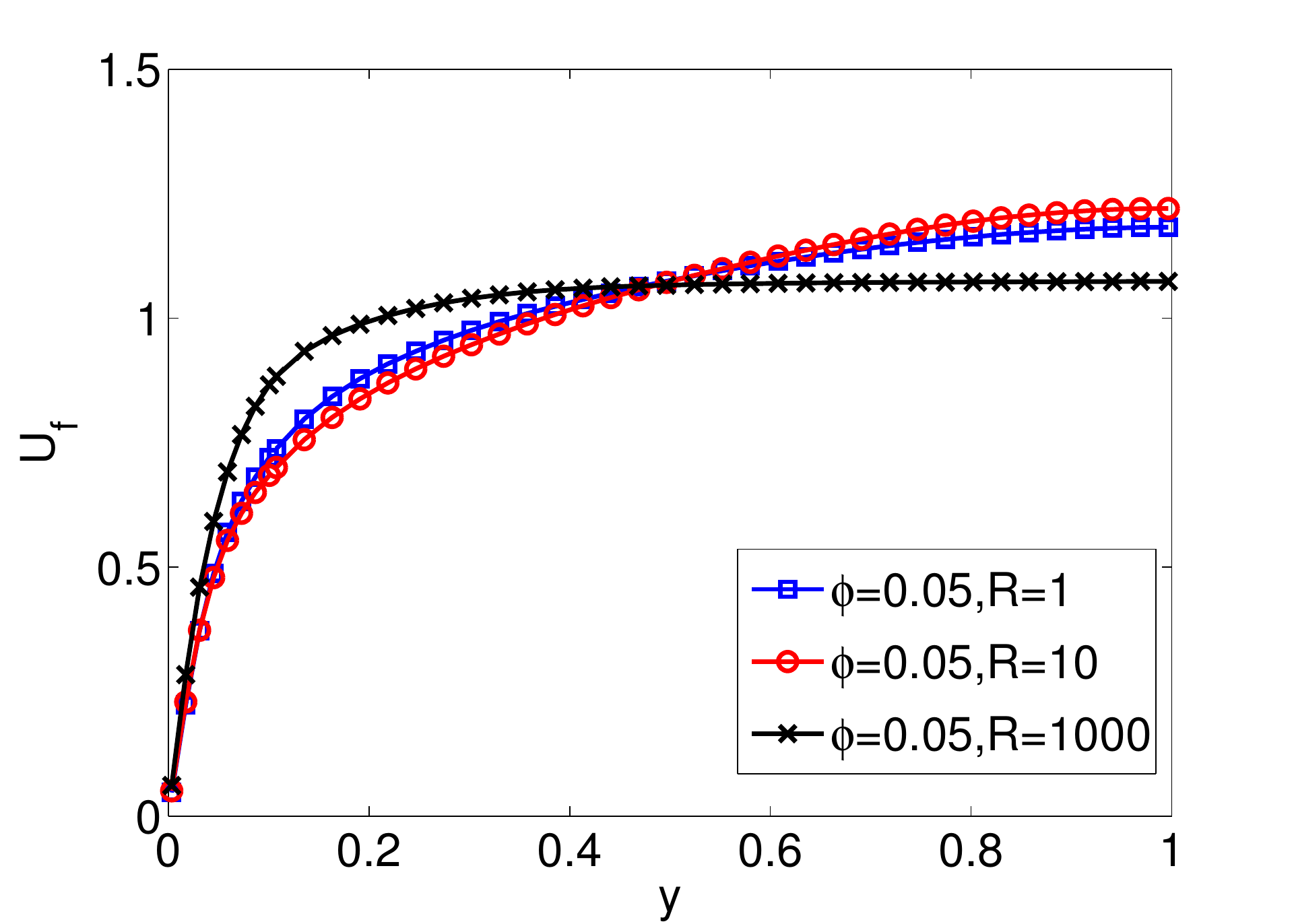}~~~~
\put(-205,110){{\large a)}}
{\includegraphics[width=.50\textwidth]{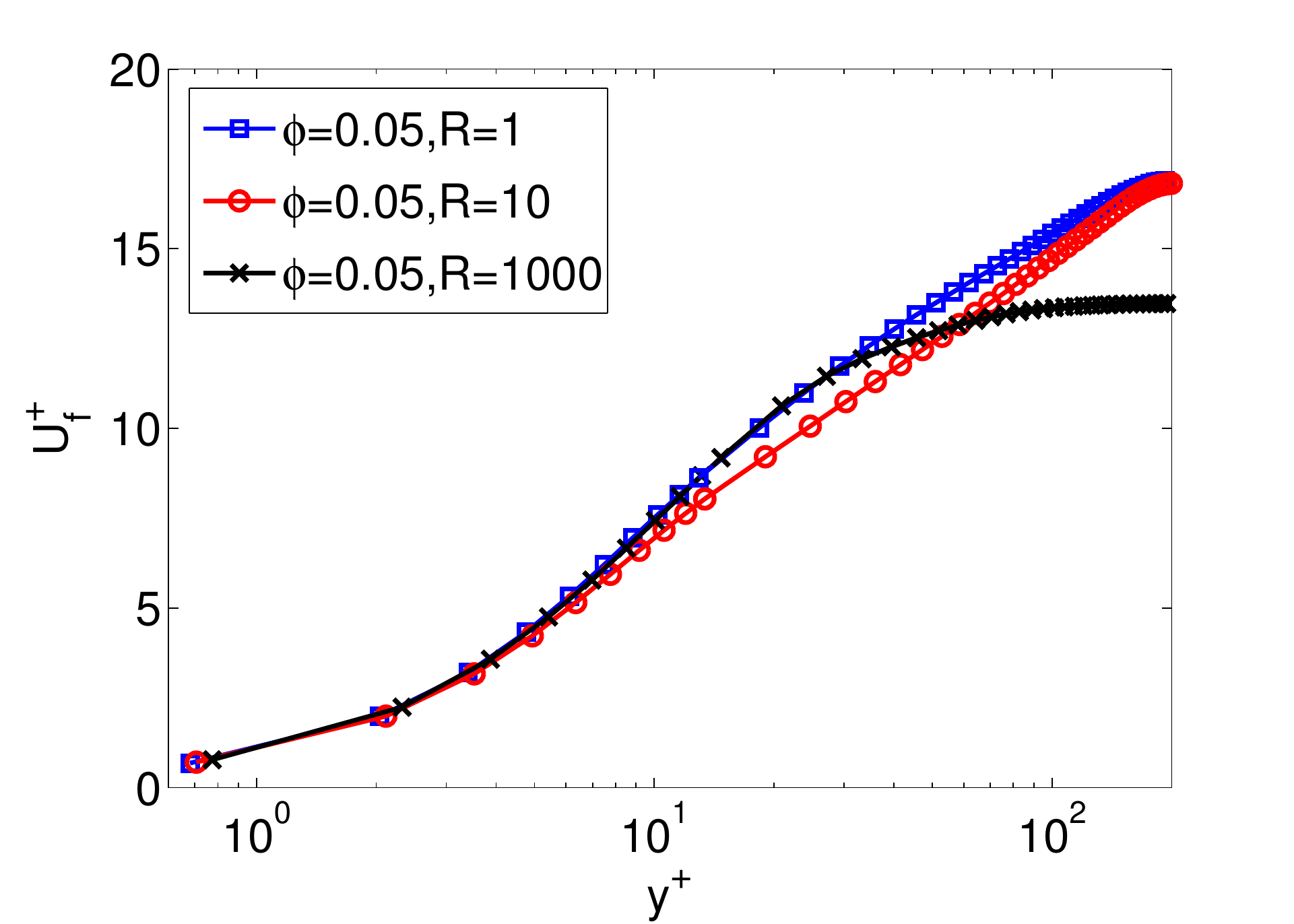}
\put(-205,110){{\large b)}}}
\includegraphics[width=.50\textwidth]{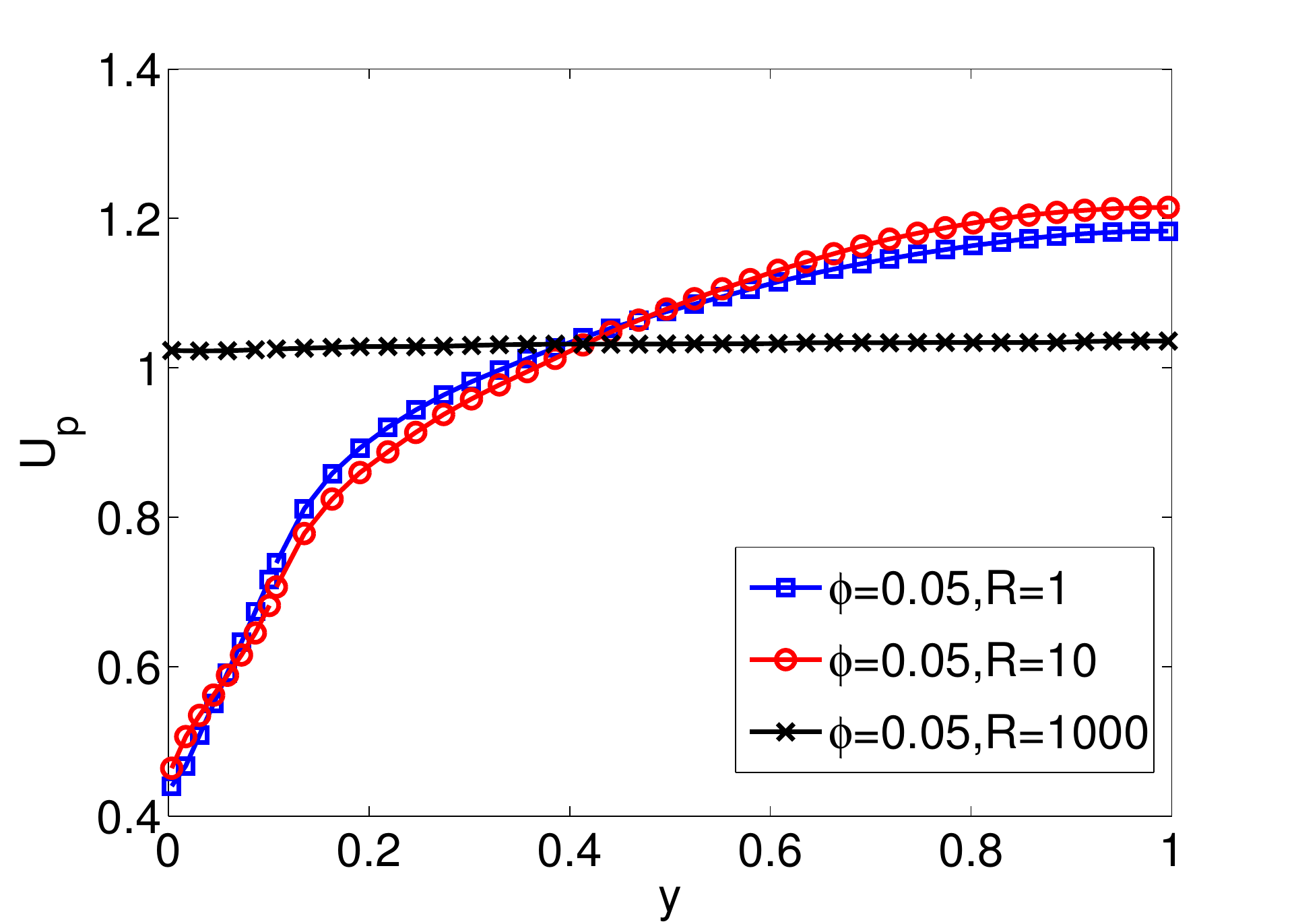}~~~~
\put(-205,110){{\large c)}}
{\includegraphics[width=.50\textwidth]{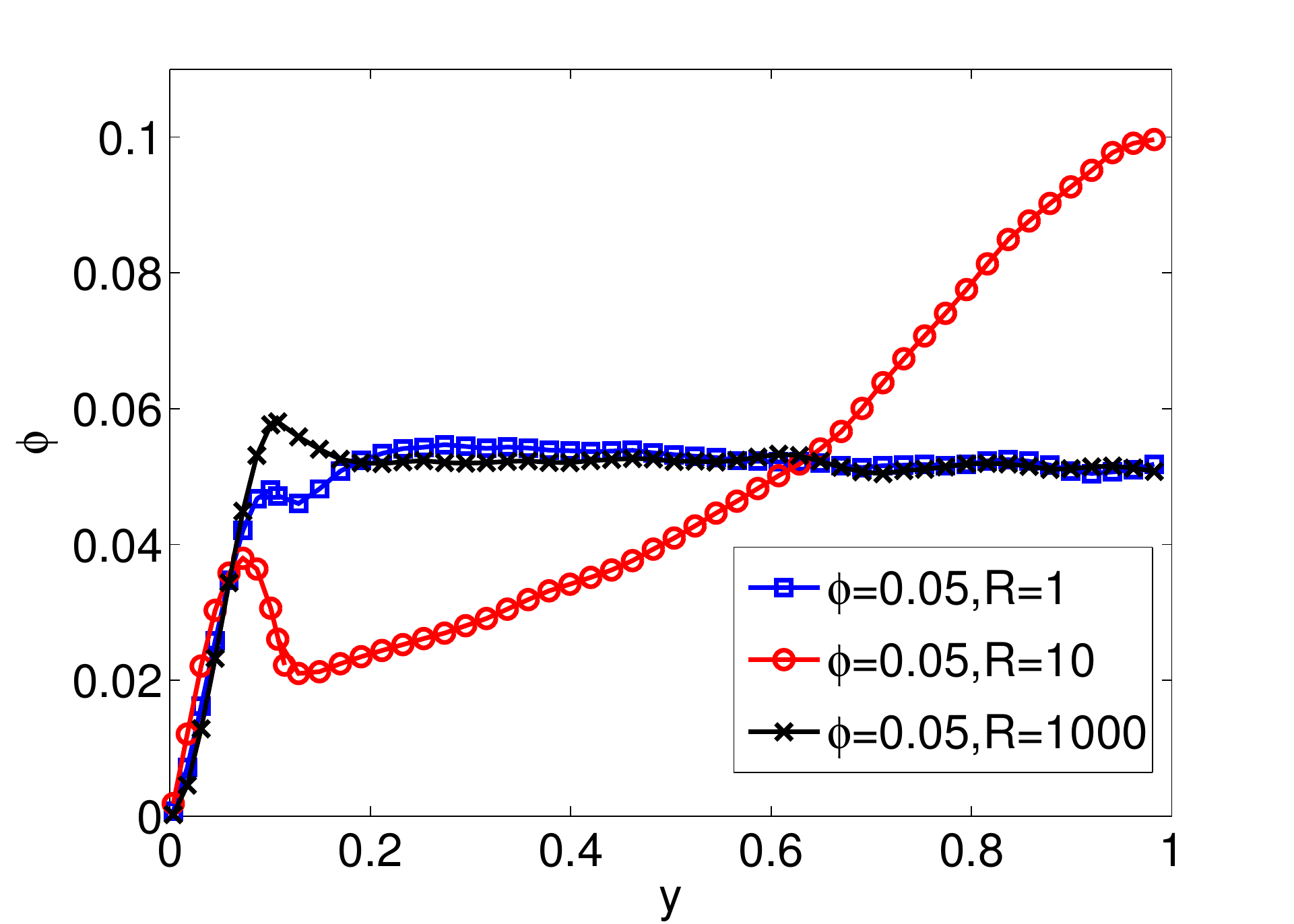}
\put(-205,110){{\large d)}}}
\caption{Panel (a) and (b): mean fluid streamwise velocity profiles in outer and inner units at constant $\phi=0.05$ and increasing $R$. Panel (c): 
mean particle streamwise velocity profiles in outer units. Panel (d): mean local volume fraction $\phi$ versus the wall-normal coordinate $y$.}
\label{fig:tutto100}
\end{figure}

\begin{figure}
\centering
\includegraphics[width=.50\textwidth]{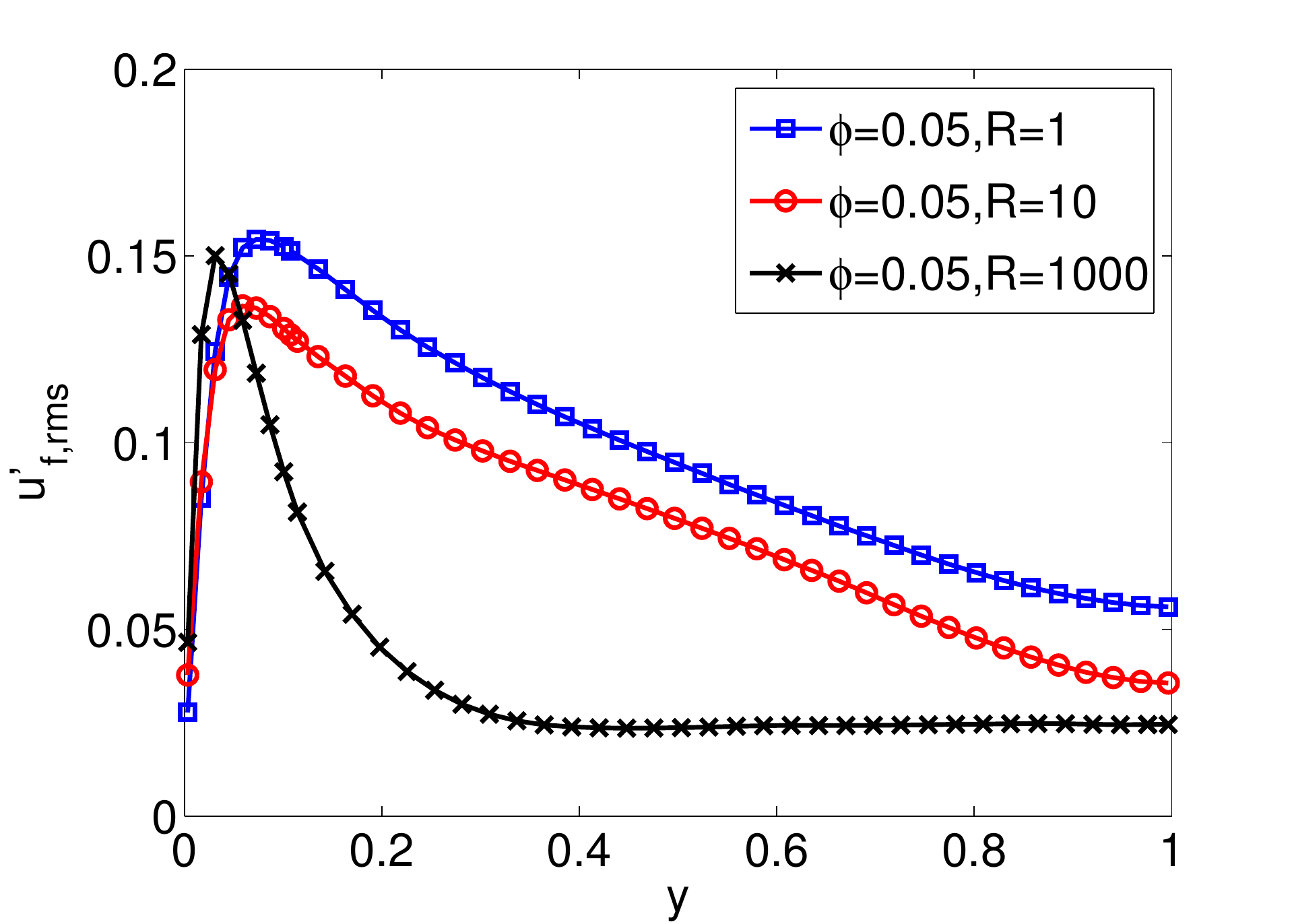}~~~~
\put(-205,110){{\large a)}}
{\includegraphics[width=.50\textwidth]{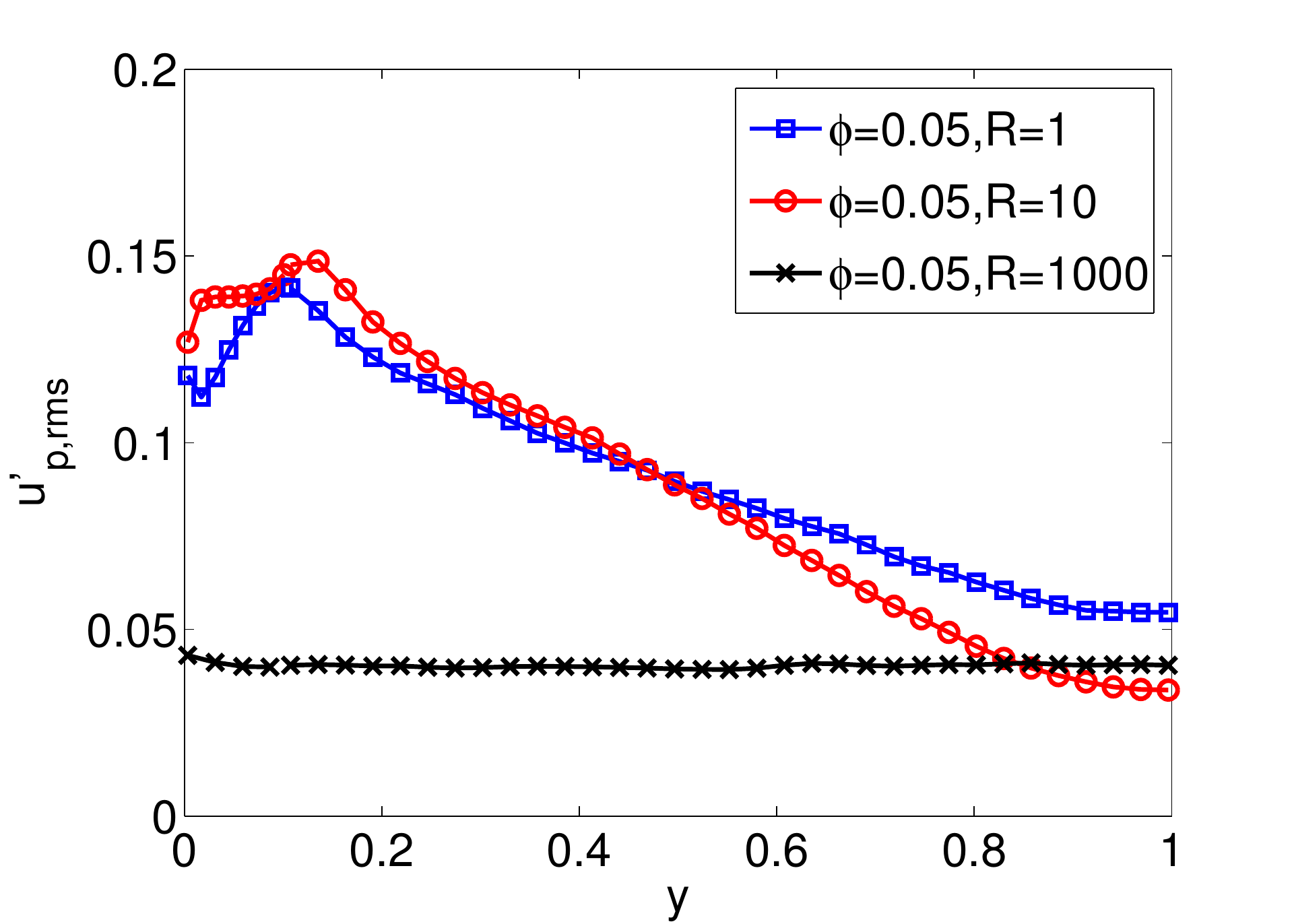}
\put(-205,110){{\large b)}}}
\includegraphics[width=.50\textwidth]{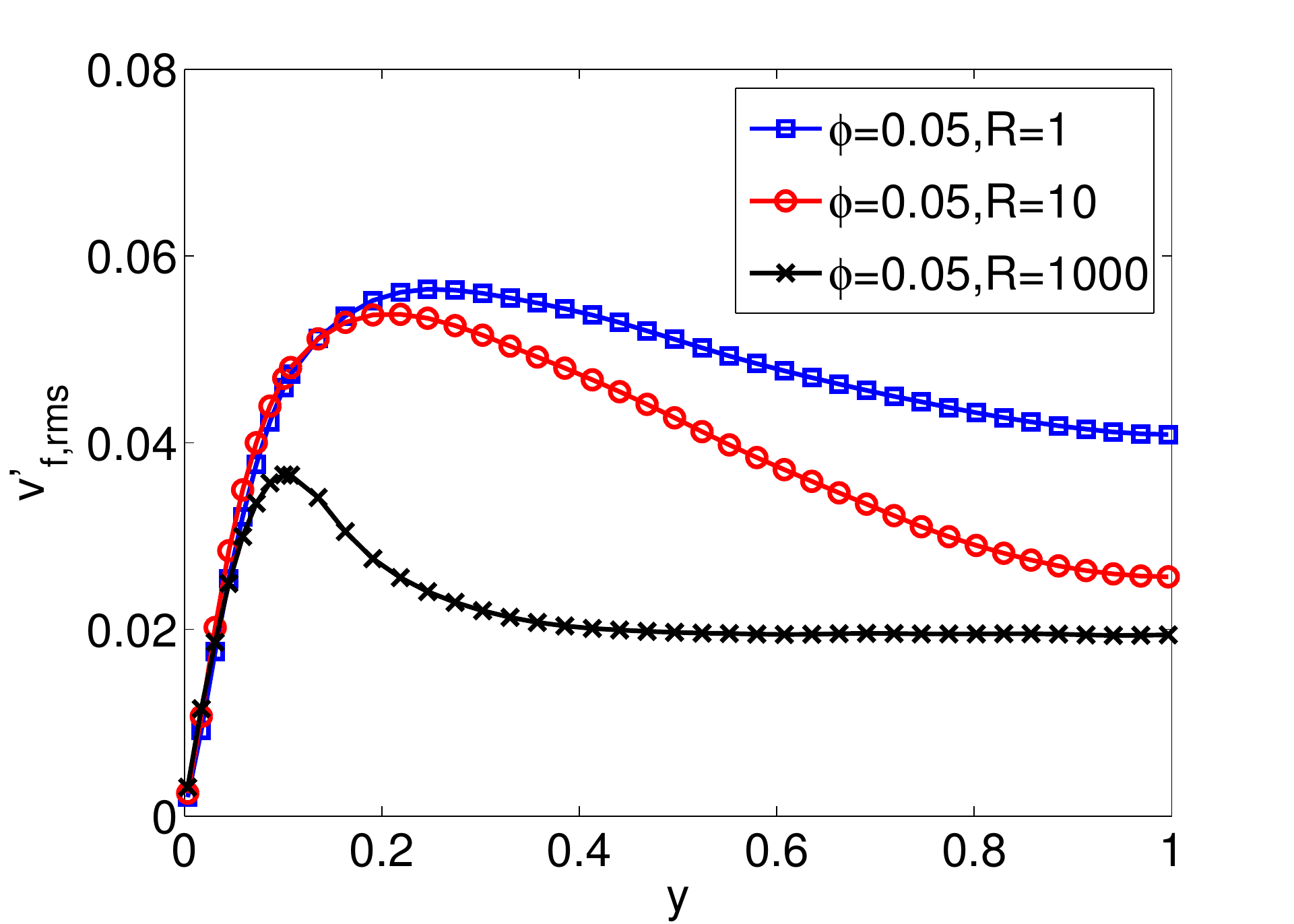}~~~~
\put(-205,110){{\large c)}}
{\includegraphics[width=.50\textwidth]{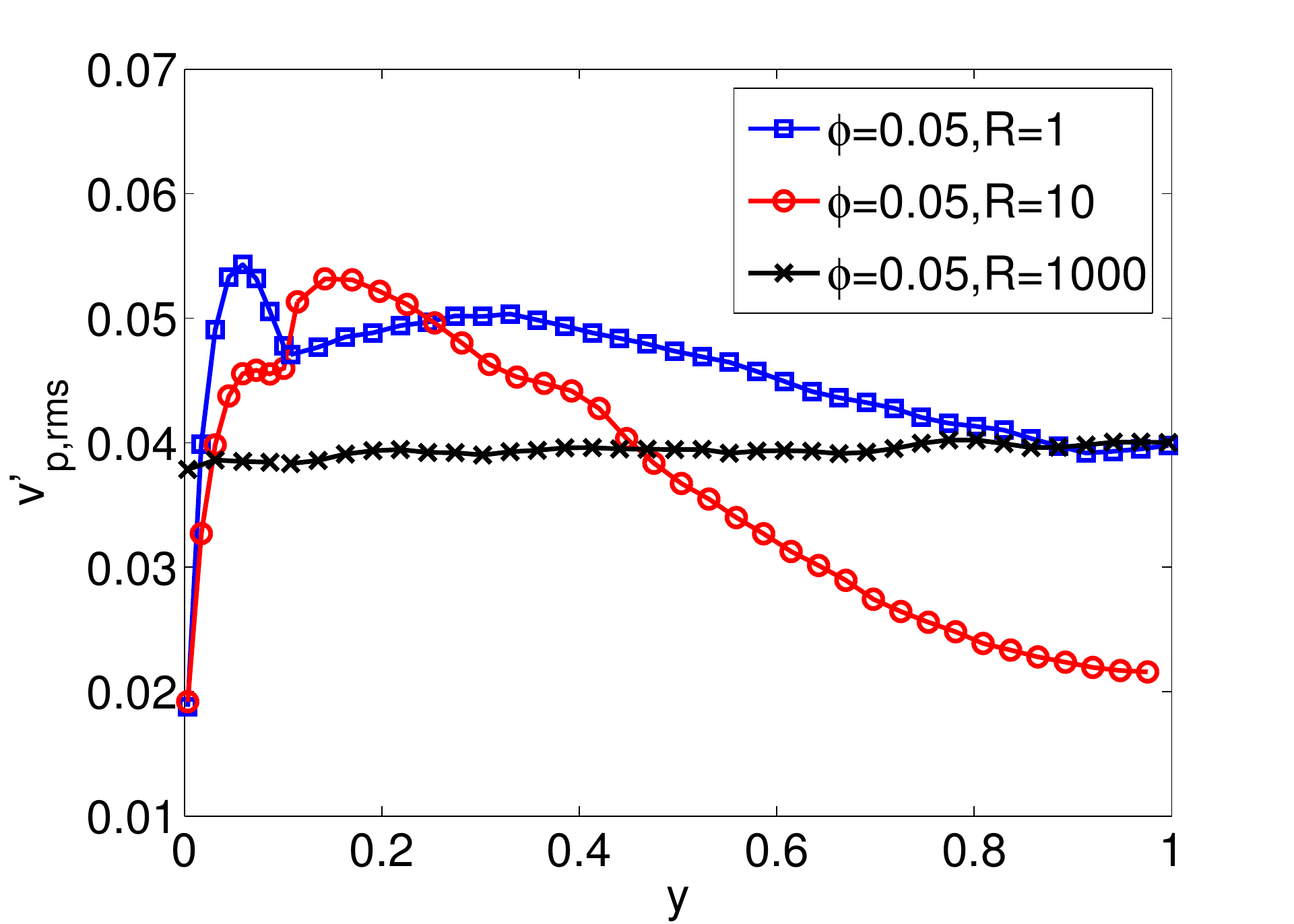}
\put(-205,110){{\large d)}}}
\includegraphics[width=.50\textwidth]{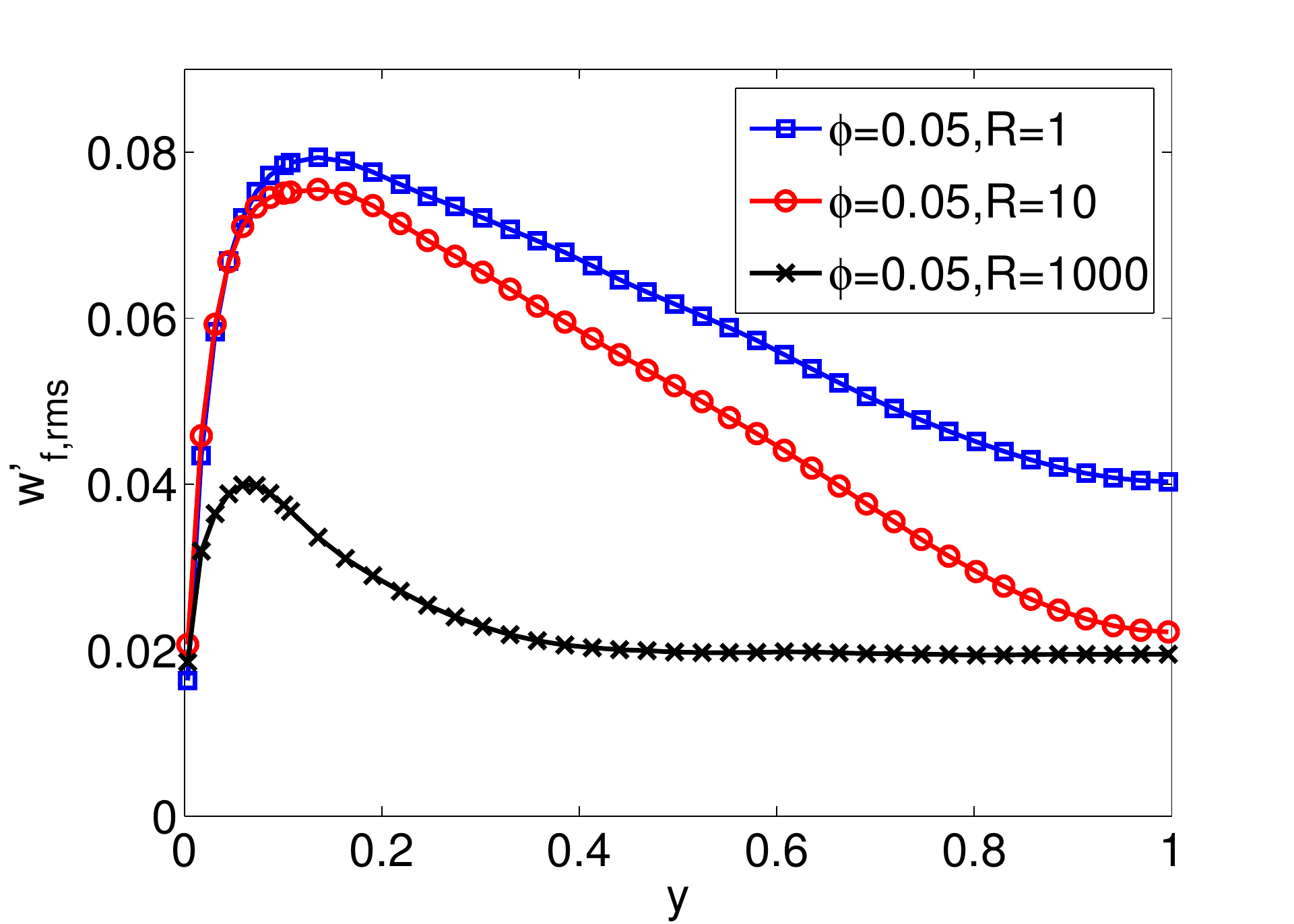}~~~~
\put(-205,110){{\large e)}}
{\includegraphics[width=.50\textwidth]{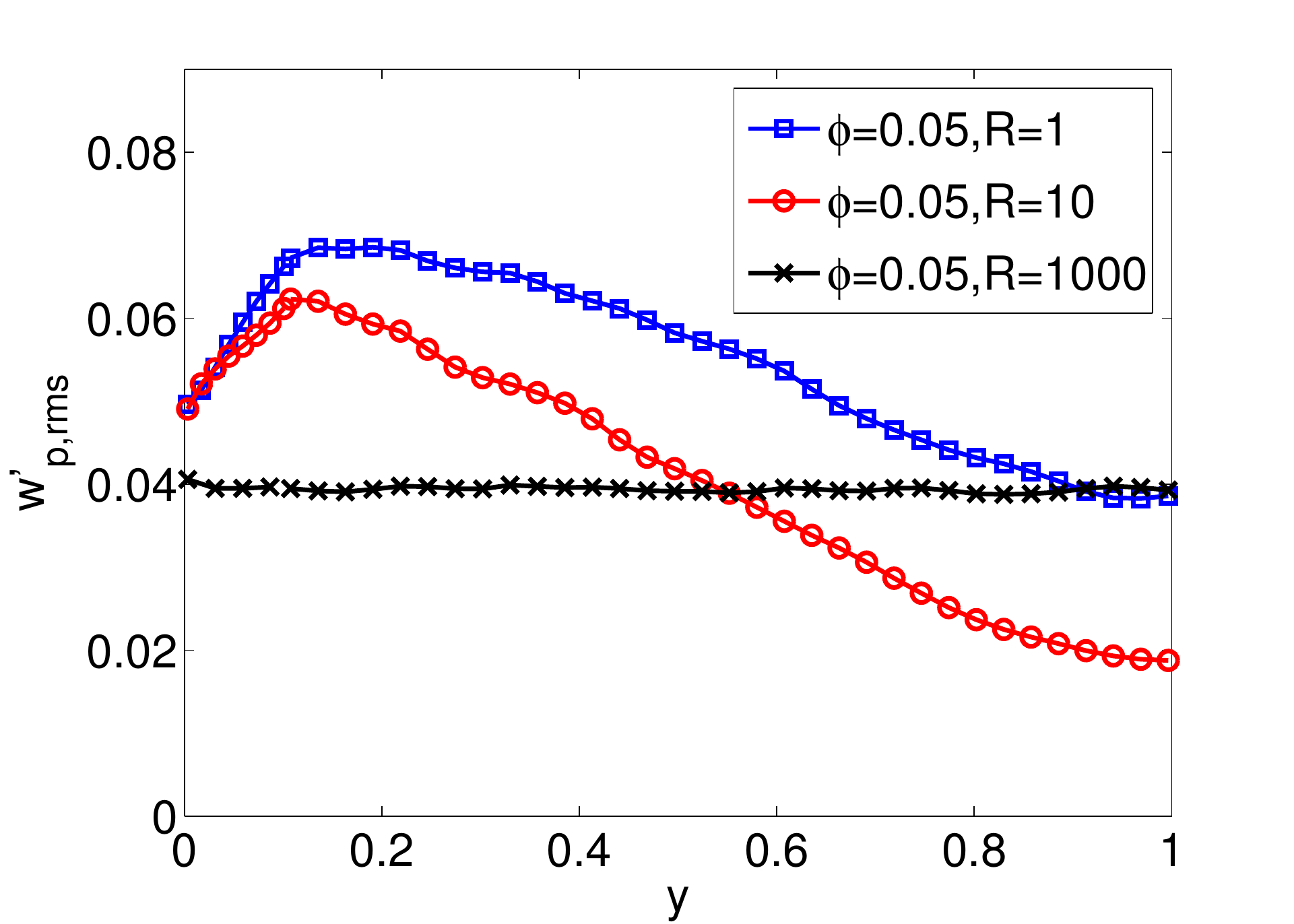}
\put(-205,110){{\large f)}}}
\caption{Intensity of the different components of the fluctuation velocities for the fluid (panels a, c and e) and for the solid phase (panels b, d  and f). the data are displayed in outer units at constant $\phi=0.05$ and increasing $R$.}
\label{fig:primi100}
\end{figure}

\subsection{Effects of Density Ratio $R$}

In this section we discuss the results obtained in an idealized scenario where the density ratio is allowed to further increase while gravity effects 
are neglected. We compare results obtained at $\phi=5\%$ and $R=1, 10$ and $1000$ showing that above a certain density ratio ($R > 10$), the 
solid phase decouples from the fluid leading to a completely different scenario. We have also looked at an intermediate case with density ratio 
$R=100$, but being the results closer to those obtained for $R=1000$, these have not been shown for sake of clarity.

\subsubsection{Single-point statistics}

The streamwise fluid velocity profiles in outer and inner units (panels (a) and (b)), the particle 
streamwise velocity profile (panel (c)) and the local volume fraction profile $\phi(y)$ (panel (d)) 
are displayed in figure~\ref{fig:tutto100}  
for $\phi=5\%$ and increasing particle density. 

The mean fluid and particle velocity $U_p$ 
changes significantly at the highest density ratio considered, $R=1000$. 
The fluid velocity increases more rapidly from the wall and reaches a constant value slight above $1$ for $y \gtrsim 0.3$. This value is about $12\%$ 
smaller than what found at the centerline for the cases with lower density ratio. 
The difference between the different profiles is even more evident when the data are scaled with inner 
units (fig.~\ref{fig:tutto100}b). 
As already mentioned in the previous section, the mean velocity profiles are similar for density ratios between 1 and 10, still giving different coefficients for the fitting of the
log-law. 
The velocity profiles almost overlap 
in the viscous sublayer
and converge to approximately the same values of $U_f^+$ for $y^+ > 100$. For the case with $R=1000$, instead, the mean velocity is close to that for $R=10$ only close to the wall, $y^+ \lesssim 20$.

Larger differences are found for the solid phase velocity, figure~\ref{fig:tutto100}(c): 
the average streamwise particle velocity is constant and approximately equal to $1$, the bulk value. 
This is similar to the behavior  previously reported for $\phi=0.2\%$ 
and $R=100$. 
All particles move in average with the same streamwise velocity, no matter if they 
are close to the walls or to the centerline. Their motion seems not to be affected by turbulent fluid flow (there is a one-way coupling between phases, since the fluid flow 
is actually modified by the presence of particles).
A pseudo-plug flow is generated across the channel, as confirmed by the local volume fraction profile, $\phi(y)$, shown in figure~\ref{fig:tutto100}(d). 
Indeed, the particles are distributed almost  uniformly across the channel, with the first particle layer appearing at approximately $2$ particle radii from the walls.

As discussed in the previous section, particle inertia and near-wall shear induce particle migration toward 
the centerline when $R=10$. 
This effect becomes more evident as the density of the particles increases, until for very high $R$, their inertia is so high 
that their motion almost completely decouples from the one of the fluid phase. In this \emph{granular-like} regime particles move ballistically between successive collisions and almost uniformly 
downstream with also an uniform wall-normal distribution.
The turbulent flow structures 
are disrupted by these heavy particles and the typical features of a turbulent channel flow are lost.

It is now interesting to look at the particle Stokes number $St_p$, the ratio between the particle time scale, due to the particle inertia, 
and a characteristic flow time scale.  
We consider the convective time as  flow characteristic  time, 
$\tau_f=h/U_0=2h^2/(Re_b \nu)$,
while the particle relaxation time is $\tau_p=\frac{4a^2 R}{18 \nu}$. 
The effect of finite inertia (i.e. of a non negligible Reynolds number) should be taken into account in the definition 
of the particle Stokes number and we therefore consider the following correction of the particle drag coefficient $C_D$ to account for inertial effects \cite{schil1935}
\begin{equation}
C_D = \frac{24}{Re_p} \left(1 + 0.15 Re_p^{0.687} \right)
\end{equation}
(where $Re_p$ is the particle Reynolds number) so that the modified Stokes number
\begin{equation}
St_p' = \frac{\tau_p}{\tau_f} \frac{1}{\left(1 + 0.15 Re_p^{0.687} \right)} = \left( \frac{2a}{h}\right)^2 \frac{1}{36} Re_b R \frac{1}{\left(1 + 0.15 Re_p^{0.687}\right)}.
\end{equation}

For sake of simplicity and in first approximation we define a shear-rate based particle Reynolds number $Re_p = Re_b (a/h)^2 \sim 20$. The modified 
Stokes number $St_p'$ then becomes equal to $0.9, 8.8$ and $883$ for $R=1, 10$ and $1000$. As expected, particle inertia becomes more and more relevant 
as the density ratio increases. For $R \in [1,10]$ the inertia of the fluid and solid phases is comparable and they mutually influence each other. 
When $R=1000$, conversely,  the particle Stokes number is much larger than 1 and particles are only slightly affected by the fluid phase. The solid phase 
behaves as a dense gas, uncorrelated to the fluid phase.

In figure~\ref{fig:primi100} we compare the fluid and particle velocity fluctuations for the three different density ratios under investigation. 
It can be seen from the plots in \ref{fig:primi100}(a),(c) 
and (e) that the fluid velocity fluctuations are significantly different at the highest $R$.
All velocity components display larger values close to the wall and then drop rapidly to a constant value of approximately $0.02$. 
Anisotropy in the energy distribution is maintained very close to the 
walls only,  whereas a quasi-isotropic energy distribution is found in the rest of the channel. The particle velocity fluctuations reported in panels (b),(d),(f) 
also exhibit an almost isotropic distribution, with a fluctuation intensity of about $0.04$.
This statistical isotropy is typical of gaseous systems and due to the strong influence of the solid phase on the fluid phase (previously explained 
by means of the particle Stokes number), the fluid velocity fluctuations are forced to approach a quasi-isotropic statistical 
steady state.

We finally observe that, approaching the centerline, particle and fluid velocity fluctuations pertaining the case with $R=10$  are smaller than those
of the neutrally buoyant case. On average, particles are more likely to be at the channel centre and move
 in the direction of the pressure gradient. Fluctuations, in all directions, are therefore reduced, and due to the strong coupling 
between the two phases, the fluid velocity fluctuations also decrease in this more ordered structure.

\subsubsection{Particle Dispersion}
 
\begin{figure}
\centering
\includegraphics[width=.50\textwidth]{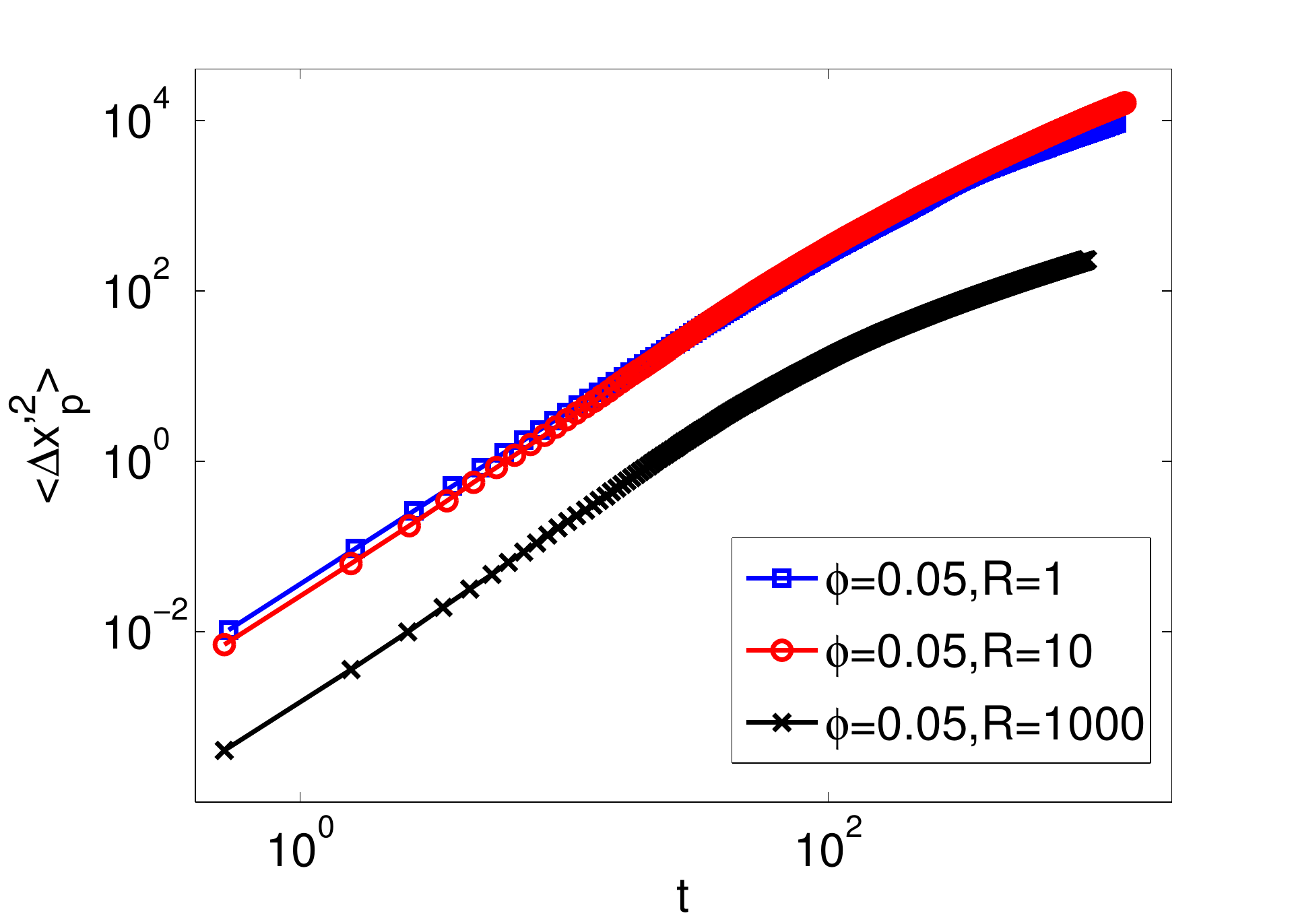}~~~~
\put(-205,110){{\large a)}}
{\includegraphics[width=.50\textwidth]{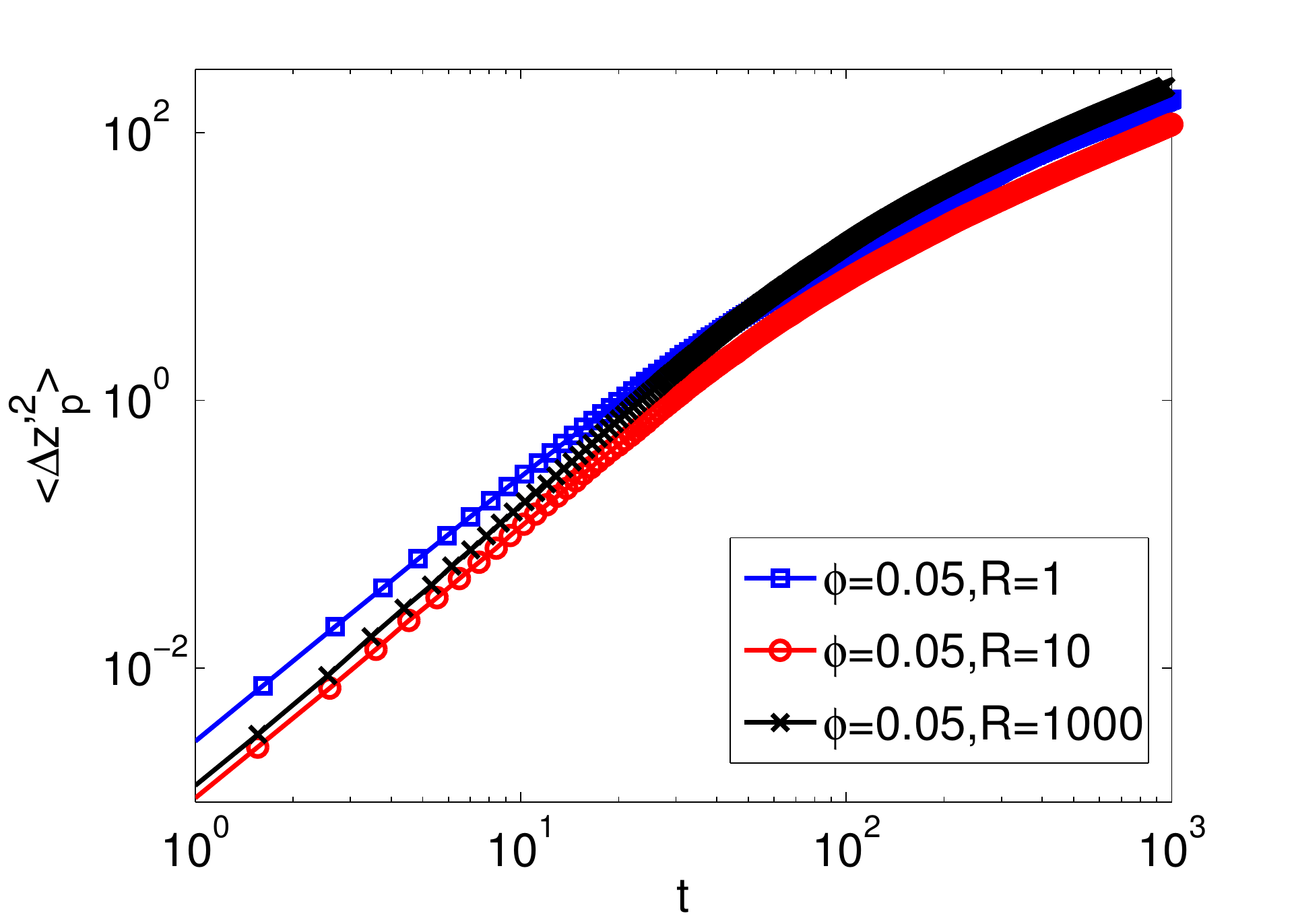}
\put(-205,110){{\large b)}}}
\caption{Particle dispersion. Time evolution of the mean square displacement along particle trajectories from the simulations at constant $\phi=0.05$ and increasing $R$. (a) Streamwise and (b) spanwise component of the dispersion.}
\label{fig:disp100}
\end{figure}

Next, we discuss the particle dispersion in the streamwise and spanwise directions. The motion of the particles is constrained in the wall-normal direction 
by the presence of the walls and is therefore not examined here.
The dispersion is quantified by the variance of the particle displacement as function of the separation time $t$. Here, we compute the mean-square displacement of the particle trajectories
\begin{equation}
\langle \Delta \vec X_p^2 \rangle(t) = \langle \left[\vec X_p(\bar t +  t) - \vec X_p(\bar t) \right]^2 \rangle_{p,\bar t}
\end{equation}
where the square displacements are averaged over time $\bar t$ and the number of particles $p$.\\
Figure~\ref{fig:disp100}(a) shows the particle dispersion in the 
streamwise direction, $\langle \Delta x_p^2 \rangle$, while the spanwise dispersion, $\langle \Delta z_p^2 \rangle$, is reported in panel (b) of the same figure.

Dispersion in the streamwise direction is similar for the cases with $R=1$ and $10$. 
The particle trajectories are initially correlated and the displacements 
proportional to time $t$. In this so-called ballistic regime, the mean square dispersion $\langle \Delta x_p^2 \rangle$ shows a quadratic 
dependence on time. Only after $t \sim 100 \, (2a)/U_0$, the curve approaches  the linear behavior typical of a diffusive motion. This is 
induced by particle-particle and hydrodynamic interactions that decorrelate the trajectories in time.

As discussed above, the motion of the solid phase is almost uncorrelated to that of the fluid when increasing the density ratio to $R=1000$. 
Since the mean particle velocity is flat across the channel, the 
dispersion is not enhanced by the inhomogeneity of the velocity profile typical of shear flows, the so-called Taylor-Aris dispersion \cite{taylor53,aris56}. 
Therefore $\langle \Delta x_p^2 \rangle$ is approximately one or two orders of magnitude lower than in the two cases at lower $R$. Interestingly, 
the purely diffusive behavior is attained faster and the transition from the ballistic behavior begins already at $t \approx 20 \, (2a)/U_0$.\\
The dispersion in the spanwise direction, $\langle \Delta z_p^2 \rangle$, is similar for all density ratios $R$ considered. 
Again, one can identify a quadratic and linear 
behavior in time with a transition between the two regimes at $t \sim 20\, (2a)/U_0$. 
We also note that, for $t \lesssim 10 \, (2a)/U_0$ 
the spanwise dispersion of the particles of highest density is close to that of particles with $R=10$, while for $t \gtrsim 200 \, (2a)/U_0$ the 
behavior appears similar to that found for $R=1$.

To conclude this section, we emphasize that the statistics of particle dispersion reveal that the particle motion only slightly changes when increasing the particle density ratio from $R=1$ to $R=10$, supporting the observation that 
the bulk flow behavior depends more on the excluded volume, i.e. $\phi$, rather than on the particle inertia.

\subsubsection{Particle velocity probability density functions and collision rates}

\begin{figure}
\centering
\includegraphics[width=.50\textwidth]{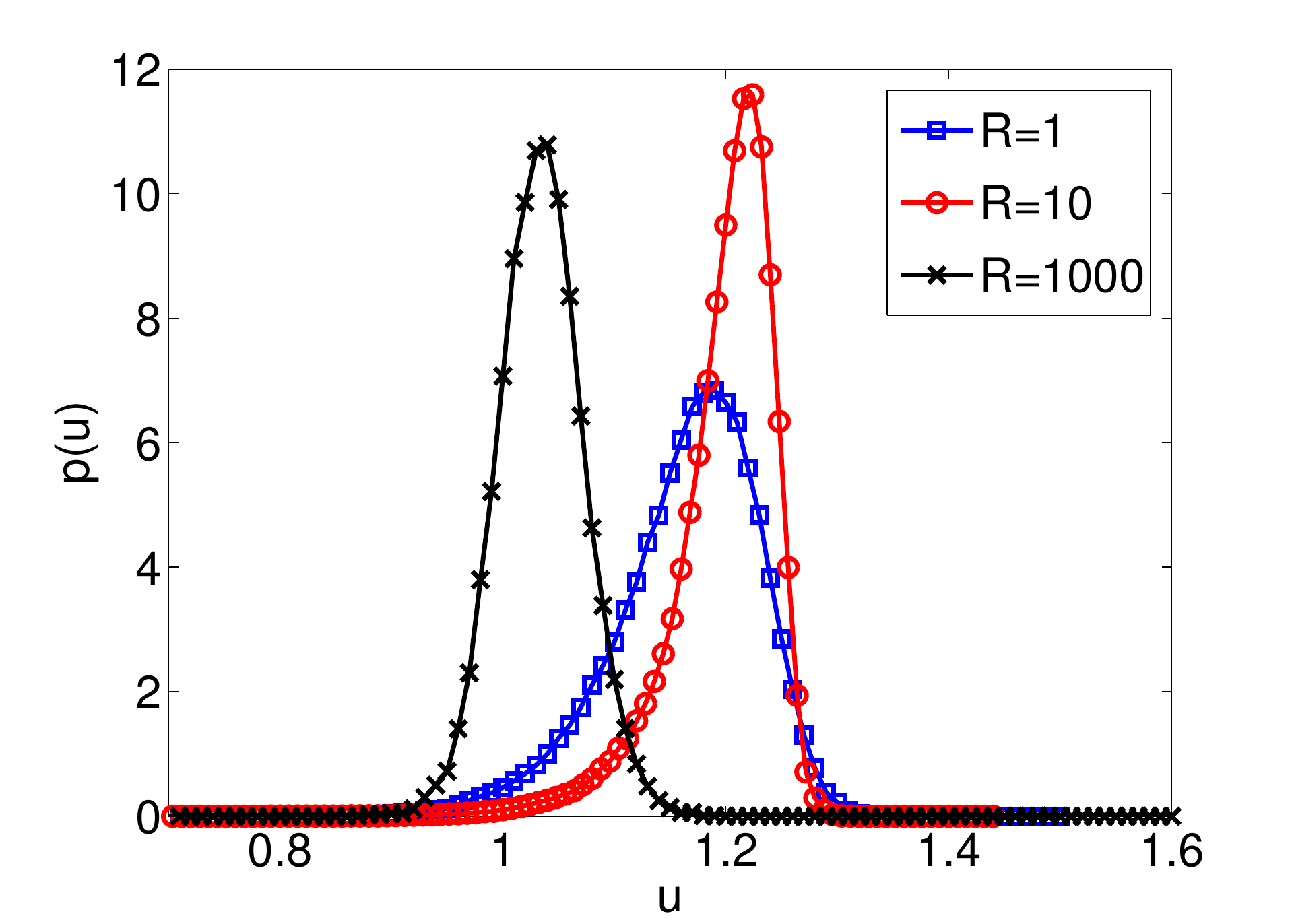}~~~~
\put(-205,110){{\large a)}}
{\includegraphics[width=.50\textwidth]{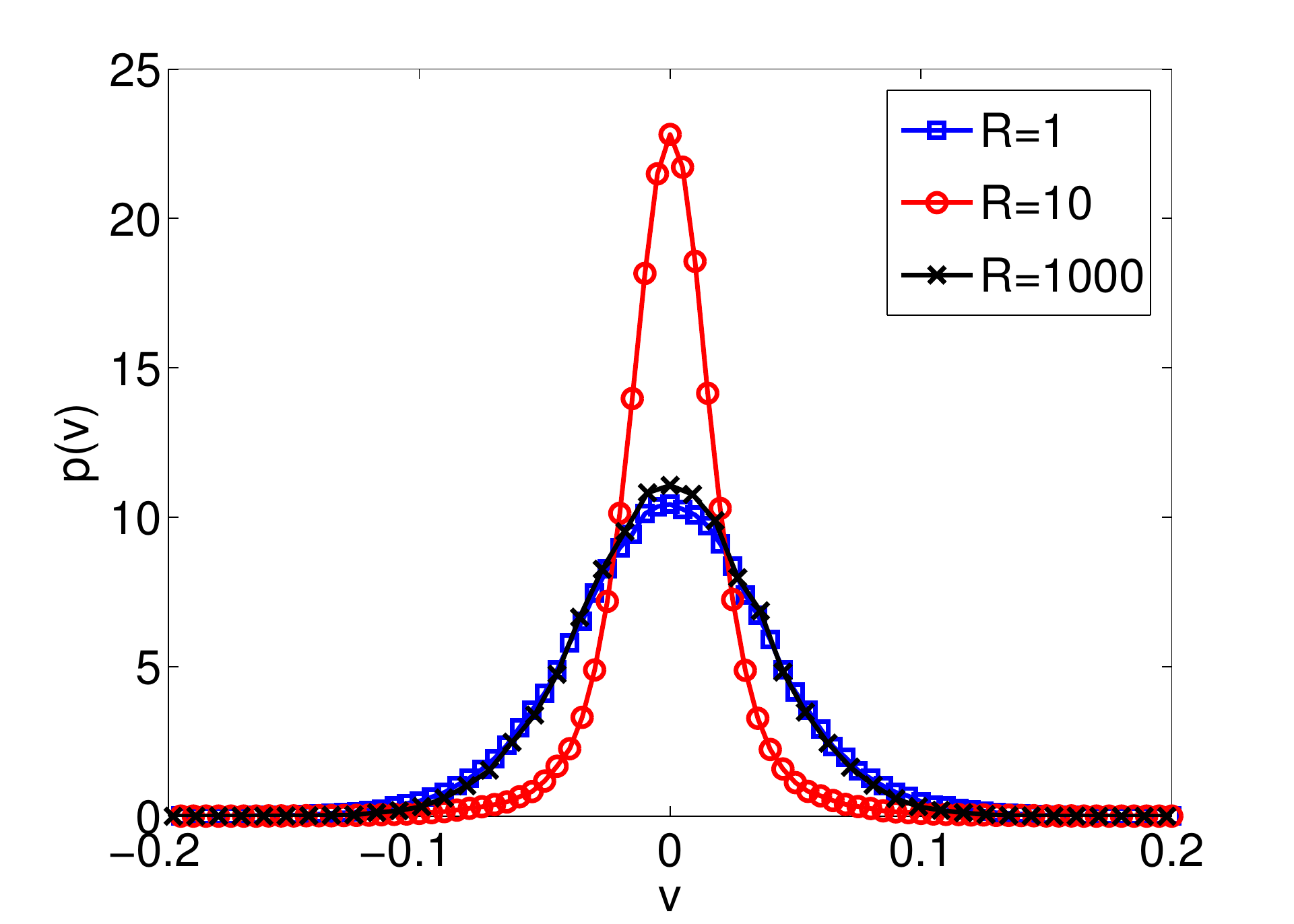}
\put(-205,110){{\large b)}}}
\includegraphics[width=.50\textwidth]{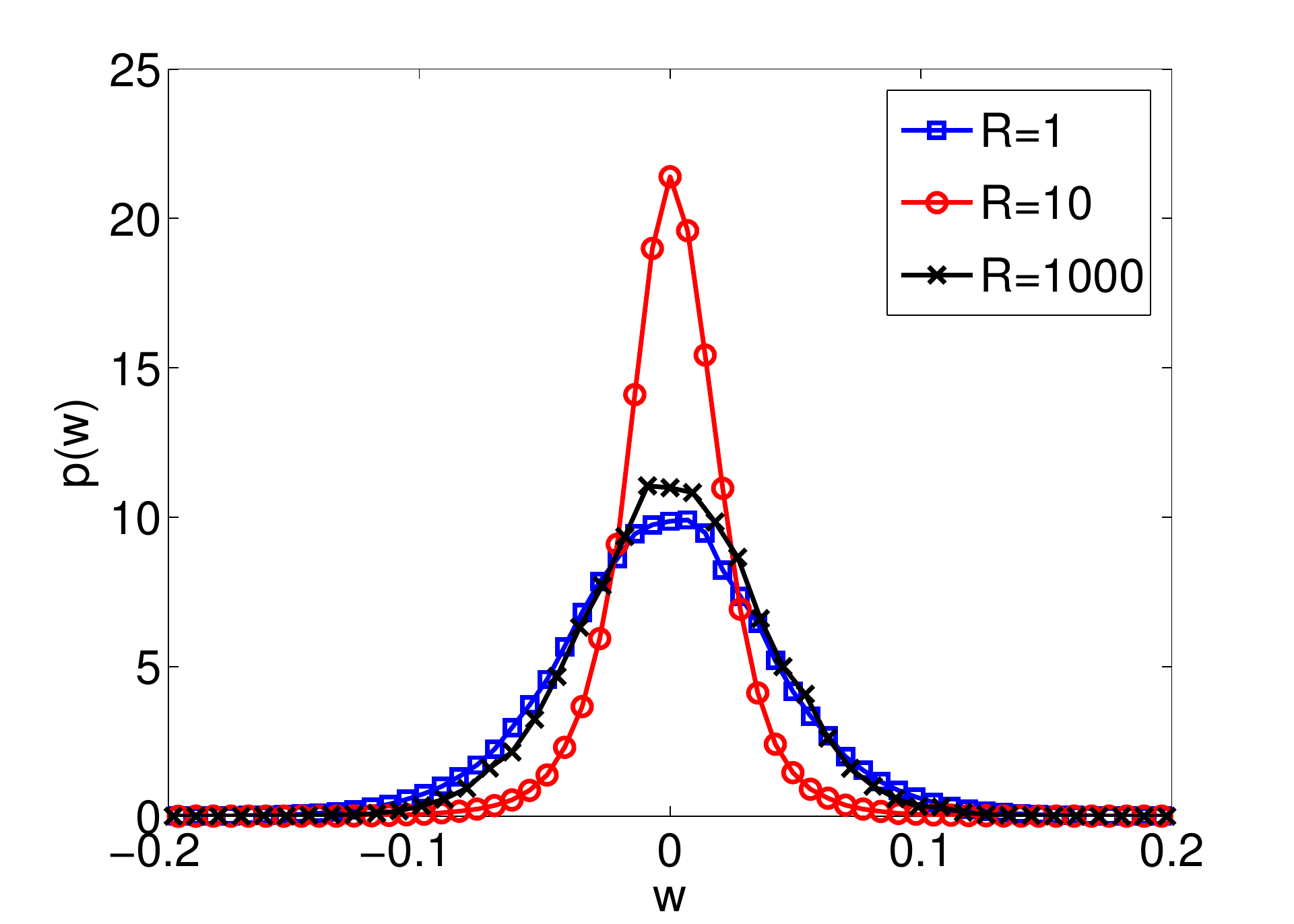}~~~~
\put(-205,110){{\large c)}}
 {\includegraphics[width=.50\textwidth]{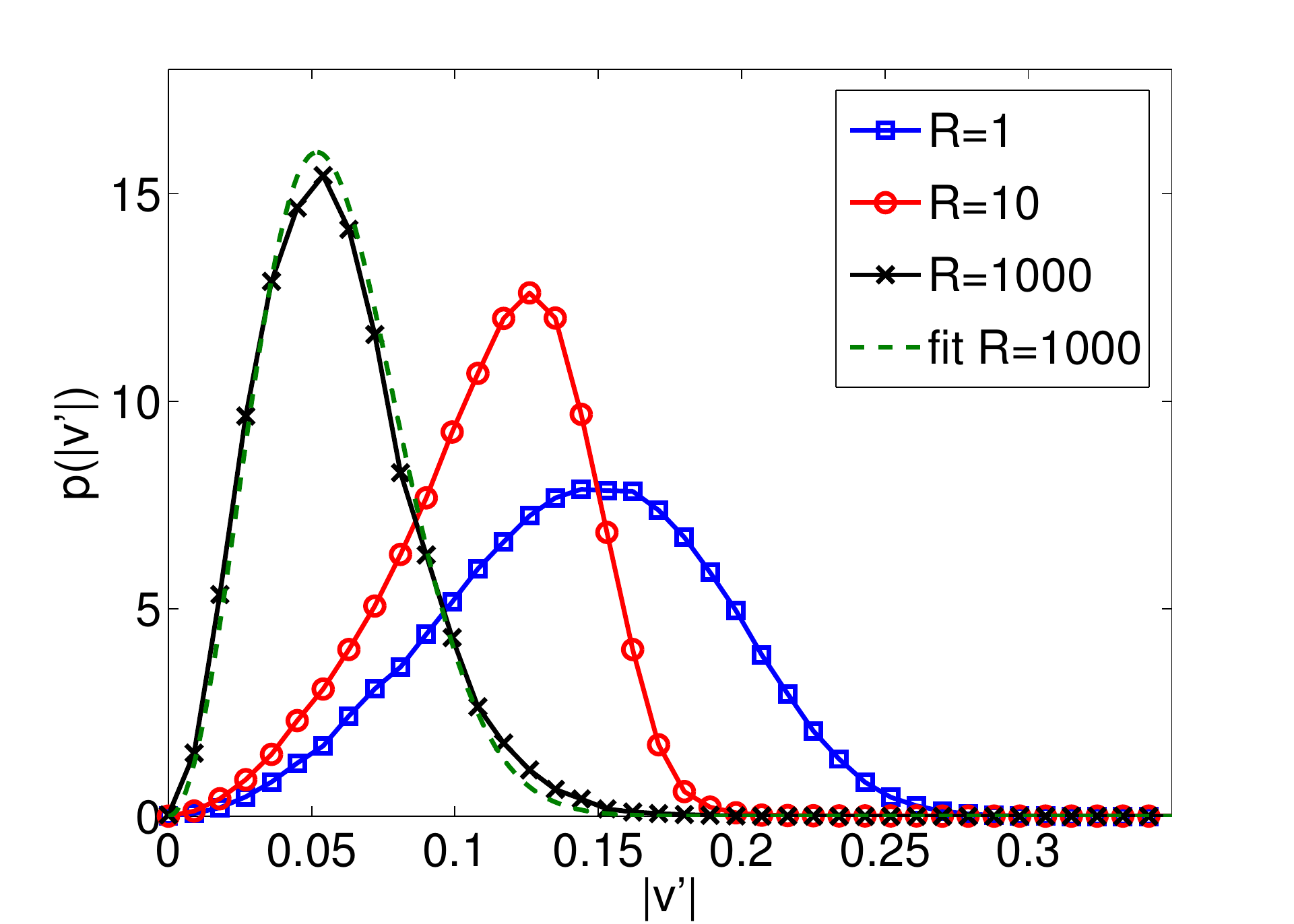}
\put(-205,110){{\large d)}}}
\caption{Probability density function of the particle velocity around the center of the channel, for the different density ratios under 
investigation (panels a, b and c for the streamwise, wall-normal and spanwise components respectively). 
Panel (d) reports the probability density function of the magnitude of the particle velocity fluctuations around the center of the 
channel where a Maxwell-Boltzmann distribution is used to fit the case with $R=1000$.}
\label{fig:pdf100}
\end{figure}

We wish to give further insight on the behavior of the solid phase dynamics by examining the velocity probability density functions. 
We will focus on the case with $R=1, 10$ and $1000$ and and calculate the probability density function $p(\cdot)$ for each component of 
the particle velocity
in the volume around the centerline of the channel (of size $2h \times 2h/3 \times 3h$). 
The distributions of the streamwise, wall-normal and spanwise 
components of the particle velocity are depicted in panels (a),(b) and (c) of figure~\ref{fig:pdf100}

We see in panel (a) that the distribution pertaining the streamwise component, $p(u)$, exhibits a negative skewness $S$ ($=-0.77$ and $-1.54$)
 for $R=1$ and $10$, indicating that particles
 exhibit  with higher probability intense fluctuations lower than the mean value, as observed also in single-phase turbulent channel flow\cite{kim1987turbulence}. 
As $R$ is increased from $1$ to $10$, the variance 
$\sigma^2$ is however reduced, whereas the flatness $F$ increases (from $3.8$ to $6.9$) indicating that rare events 
become more frequent. 
The results for $R=1000$ show that the velocity distribution changes to what, at first sight, may seem a normal distribution with smaller modal value 
and variance, almost vanishing skewness ($S \sim 0.03$) and flatness close to $3$ ($F \sim 3.3$).

The velocity distributions in the cross-stream
directions (reported in panels b and c) resemble a normal distribution centered around 
a zero mean value. As for the streamwise component, the flatness $F$ exhibits high values (between $6$ and $7$) only for $R=10$ while for the remaining 
two cases it is just slightly greater than $3$.

Next, we report the probability distributions of the modulus of the velocity fluctuations, 
\begin{equation}
|v'|=\sqrt{u_{p,rms}'^2 + v_{p,rms}'^2 + w_{p,rms}'^2},
\end{equation} 
calculated in the same volume around the centerline in figure~\ref{fig:pdf100}(d). The most peculiar distribution is the one found for $R=1000$. It closely resembles 
a Maxwell-Boltzmann distribution (or a $\chi$ distribution with three degrees of freedom) defined as follows:
\begin{equation}
\label{eqMB}
p(x) = \sqrt{\frac{2}{\pi}} \frac{x^2 e^{-x^2/(2a^2)}}{a^3},
\end{equation}
where $a$ is a scale parameter (velocity). This distribution describes the velocity of atoms of an ideal gas that freely move  inside a stationary 
container. 
In such case the scale parameter becomes $a=\sqrt{kT/m}$ where $k$ is the Boltzmann's constant, $T$ the thermodynamic temperature and 
$m$ the particle mass. Fitting our results with equation~\ref{eqMB} we find $a \sim 0.037$, corresponding to the Maxwell-Boltzmann distribution 
displayed in panel (d) with dashed line. 
The root mean square of such a distribution is $\sigma = \sqrt{3} a =0.064$, using the value of $a$ 
previously reported. 
Examining again figure~\ref{fig:primi100}(b),(d),(f), we notice that the velocity fluctuations are 
approximately equal to $0.04$, with modulus $|v'| \simeq 0.069$. Thus the root mean square $\sigma$ is completely defined by $|v'|$. 
These findings further confirm our previous speculations about the appearance of a dense gaseous regime at 
 high density ratios $R$.

Finally we examine particle-pair statistics,  function of the distance between the centers $r$, and show that the large variations of the
particle velocity also affect the particle-pair dynamics, in particular the collisions. As the distance $r$ approaches the particle diameter, the 
near field interactions become important and collisions may occur (whenever $r = 2a$). An indicator of the radial separation among pair of particles is 
the Radial Distribution Function $RDF$. In a reference frame with origin at the centre of a particle, the $RDF$ is the average number of particle 
centers located in the shell  of radius $r$ and thickness $\Delta r$, normalized with the number of particles of a random distribution. Formally 
the $RDF$ is defined as 
\begin{equation}
\label{eqrdf}
RDF(r) = \frac{1}{4 \pi} \td{N_r}{r}\frac{1}{r^2 n_0},
\end{equation}
where $N_r$ is the number of particle pairs on a sphere of radius $r$, $n_0=N_p(N_p-1)/(2V)$ is the density of particle pairs in the volume $V$, with 
$N_p$ the total number of particles. 
The value of the $RDF$ at distances of the order of the particle radius reveals the intensity of clustering; the RDF tends to $1$ as 
$r \to \infty$, corresponding to a random (Poissonian) distribution.

\begin{figure}
\centering
\includegraphics[width=.50\textwidth]{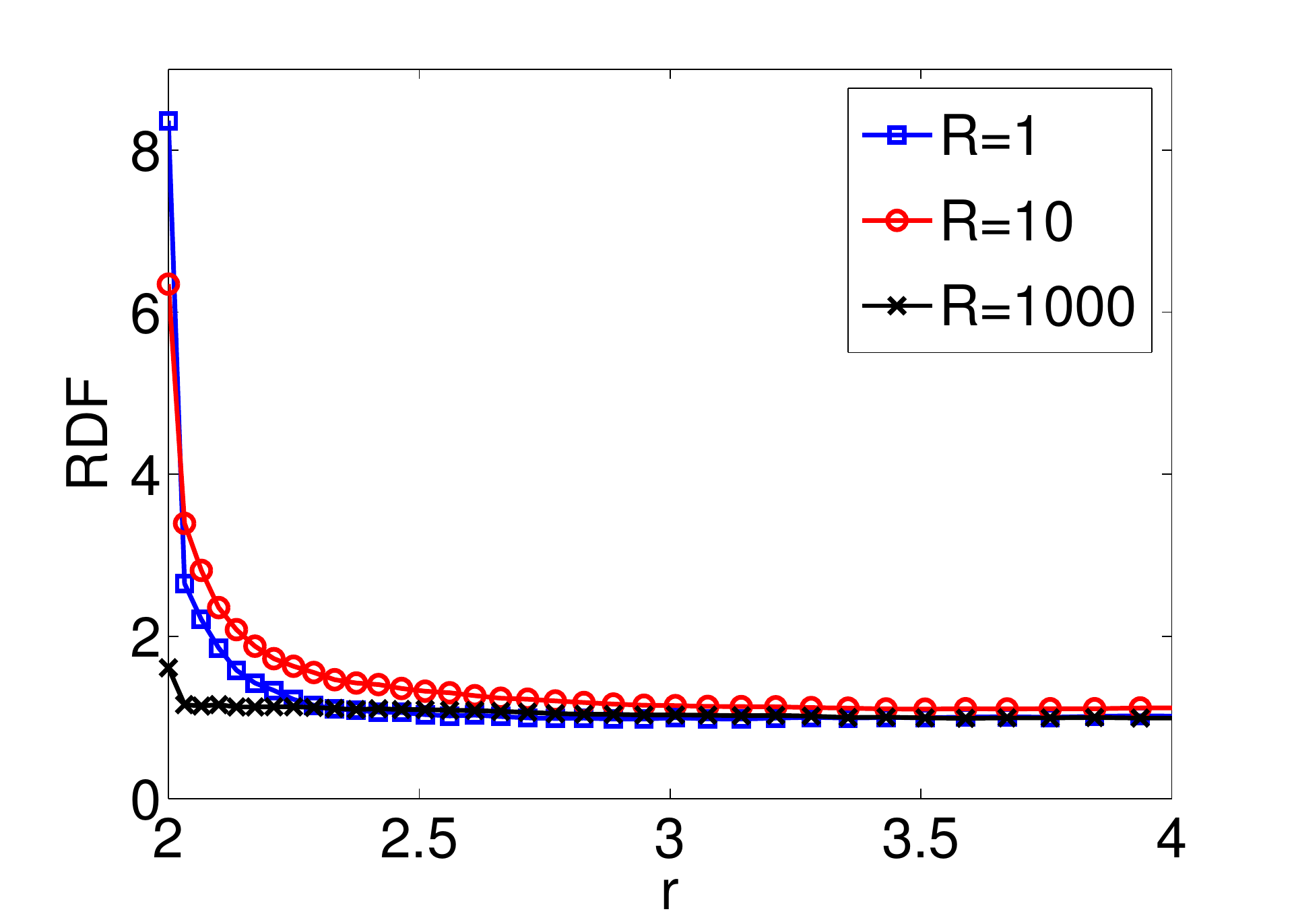}~~~~
\put(-205,110){{\large a)}}
{\includegraphics[width=.50\textwidth]{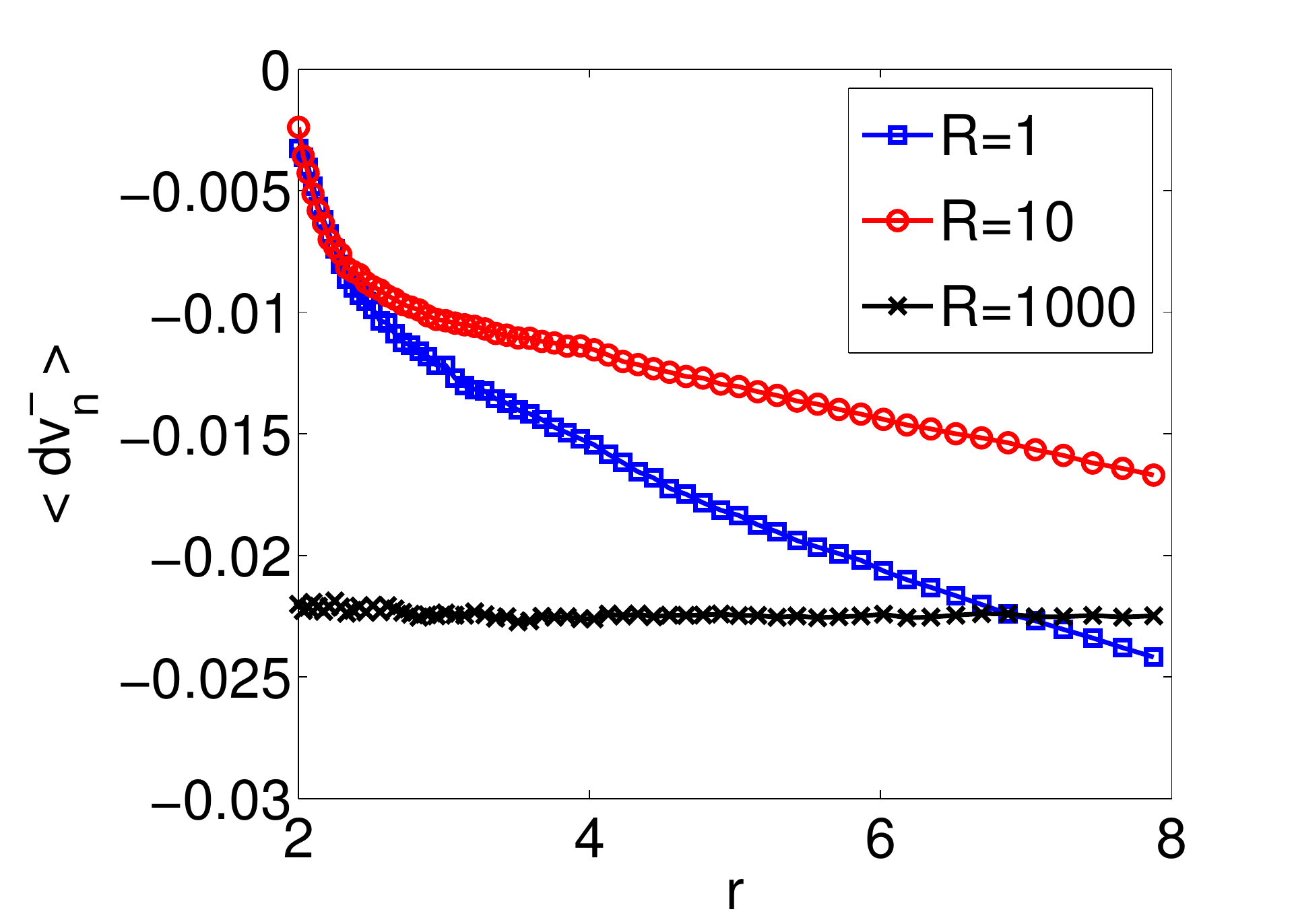}
\put(-205,110){{\large b)}}}
\includegraphics[width=.50\textwidth]{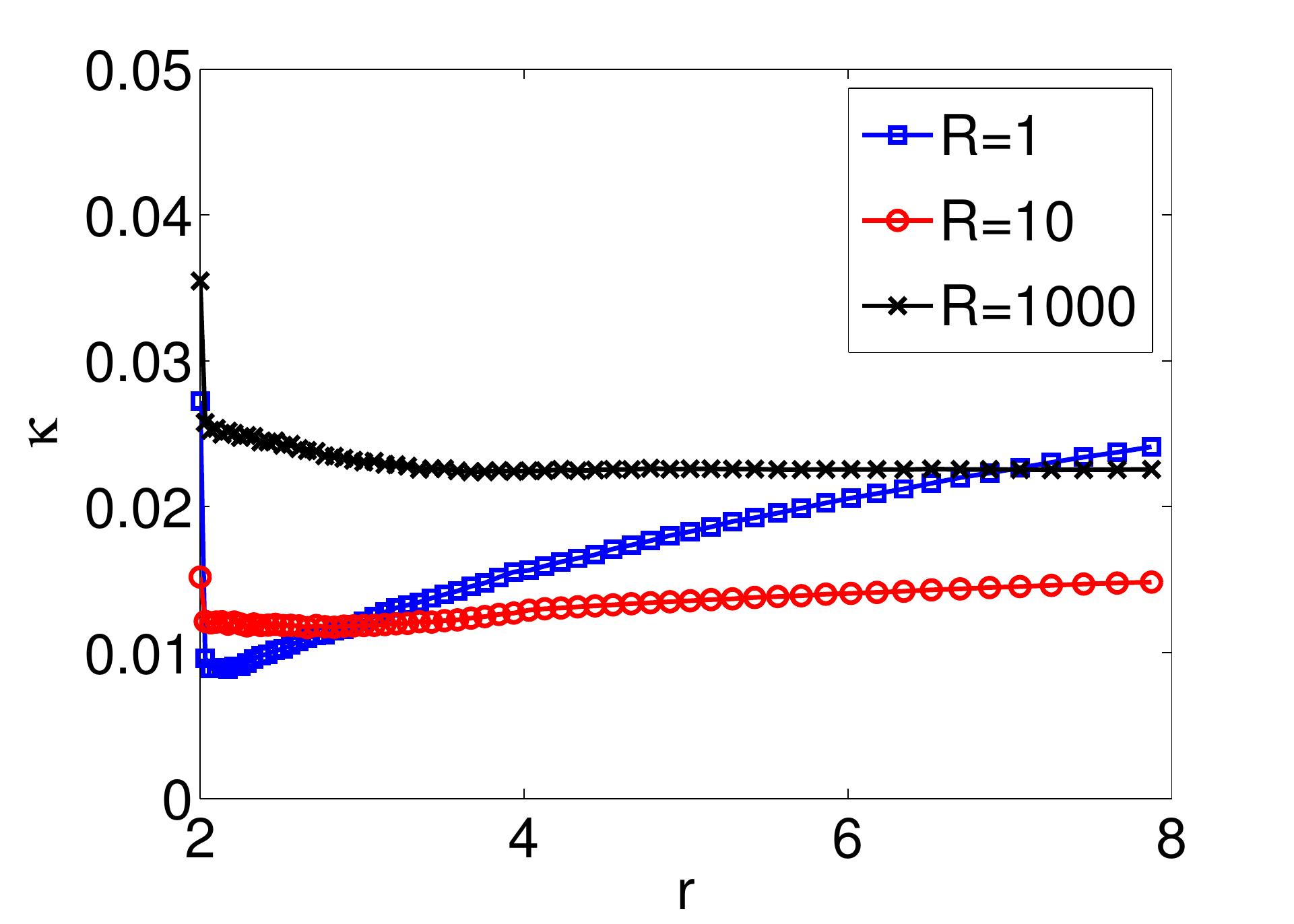}
\put(-205,110){{\large c)}}
\caption{(a) Radial distribution function, (b) average negative relative velocity and (c) collision kernel (see text for the definitions) around the centerline for the three values of the density ratio $R$ indicated and particle volume fraction $\phi=5\%$. Distances are normalized by the particle radius.}
\label{fig:rdf100}
\end{figure}

Here, we are mainly interested  in the particle-pair statistics around the centerline, 
and therefore compute the $RDF$ 
in the volume  defined by $y \in [0.67, \,1.33]$ 
for the three density ratios $R=1, 10$ and $1000$ and volume fraction $\phi=5\%$.
The data obtained are shown in figure~\ref{fig:rdf100}(a). 
At lower density ratios, $R=1$ and 10,  the peaks of the $RDF$'s are 
found at exactly $2$ particle radii from the centre of the reference particles. 
The $RDF$ drops quickly to the 
value of the uniform distribution (i.e.\ $1$) at $r \sim 2.25 a$ in the neutrally buoyant case, whereas the decay is somewhat slower for $R=10$, reaching the final plateau   at $r \sim 3$. 
This difference can be explained by the shear-induced migration 
previously discussed: this enhances the number of particles around the centerline, thus increasing the local volume fraction and consequently the small scale 
clustering. At the highest density ratio under investigation, instead, the gaseous behavior of the solid phase leads to an uncorrelated statistical distribution of 
particles, corresponding  to a constant value of the $RDF$ equal to $1$.

Figure~\ref{fig:rdf100}(b) and (c) show the averaged normal relative velocity between two  approaching particles 
$\langle dv_n^-(r) \rangle$, and the collision kernel $\kappa(r)$. This collision kernel \cite{sundaram1997} is obtained as the product of the $RDF(r)$ and 
$\langle dv_n^-(r) \rangle$: 
\begin{equation}
\kappa(r) = RDF(r) \cdot |\langle dv_n^-(r) \rangle|,
\end{equation}
when $r=2a$. In the figure, we display the behavior of this observable with the distance $r$, which can be interpreted as the approach
rate of particle pairs at distance $r$. 
The normal relative velocity of a particle pair is obtained as the projection of the relative velocity in the direction of the distance between the two interacting particles
\begin{equation}
dv_n(r_{ij}) = (\vec u_i - \vec u_j) \cdot \frac{(\vec r_i - \vec r_j)}{|(\vec r_i- \vec r_j)|} = (\vec u_i - \vec u_j) \cdot \frac{\vec r_{ij}}{|\vec r_{ij}|}
\end{equation}
(where $i$ and $j$ denote the two  particles). This scalar quantity can be either positive (when two particles depart form each other) or negative 
(when they approach). Hence, the averaged normal relative velocity can be decomposed into $\langle dv_n(r) \rangle = \langle dv_n^+(r) \rangle + \langle 
dv_n^-(r) \rangle$. To estimate the probability of a collision, i.e.\ the collision kernel $\kappa(r)$, the 
mean negative normal relative velocity is therefore needed.

It is shown in figure~\ref{fig:rdf100}(b) that the absolute value of $\langle dv_n^-(r) \rangle$ increases with $r$ when $R \le 10$. Particle pairs are more likely to approach 
with higher speeds when further away. This increase of $|\langle dv_n^-(r) \rangle|$ with $r$ is less pronounced for $R=10$, which can be explained recalling that, in this case,
there is a significant accumulation in the region around the centerline where  the particles are transported downstream at almost constant velocity. When $R=1000$, 
$|\langle dv_n^-(r) \rangle|$ is constant and equal to $0.022$. In a dense gaseous regime, particles are, on average, uniformly distributed and approach each other at 
similar speeds and at different radial locations: their motion is uncorrelated.

The collision rate is mainly determined by the averaged normal relative velocity when $R=1000$. As shown in figure~\ref{fig:rdf100}(c),
 $\kappa(r)$ is approximately constant at different radial distances, showing slightly larger values near contact, $r=2a$. In the 
cases with $R=1$ and 10,  $\kappa(r)$ is determined at small separations $r$  by the particle clustering and by the normal relative velocities at higher 
separations. When shear-induced migration occurs, $R=10$, the collision kernel $\kappa(r)$ is higher than 
in the case of neutrally buoyant particles 
for separations between $2$ and $3$ particle 
radii. When $r \gtrsim 3$ the Radial Distribution Function drops to $1$ and the approach rate is therefore determined by the averaged normal relative 
velocity. Since the absolute value of  $|\langle dv_n^-(r) \rangle|$  grows more slowly with $r$ for $R=10$, $\kappa(r)$ shows the same 
trend.

\begin{figure}
\centering
\includegraphics[width=.50\textwidth]{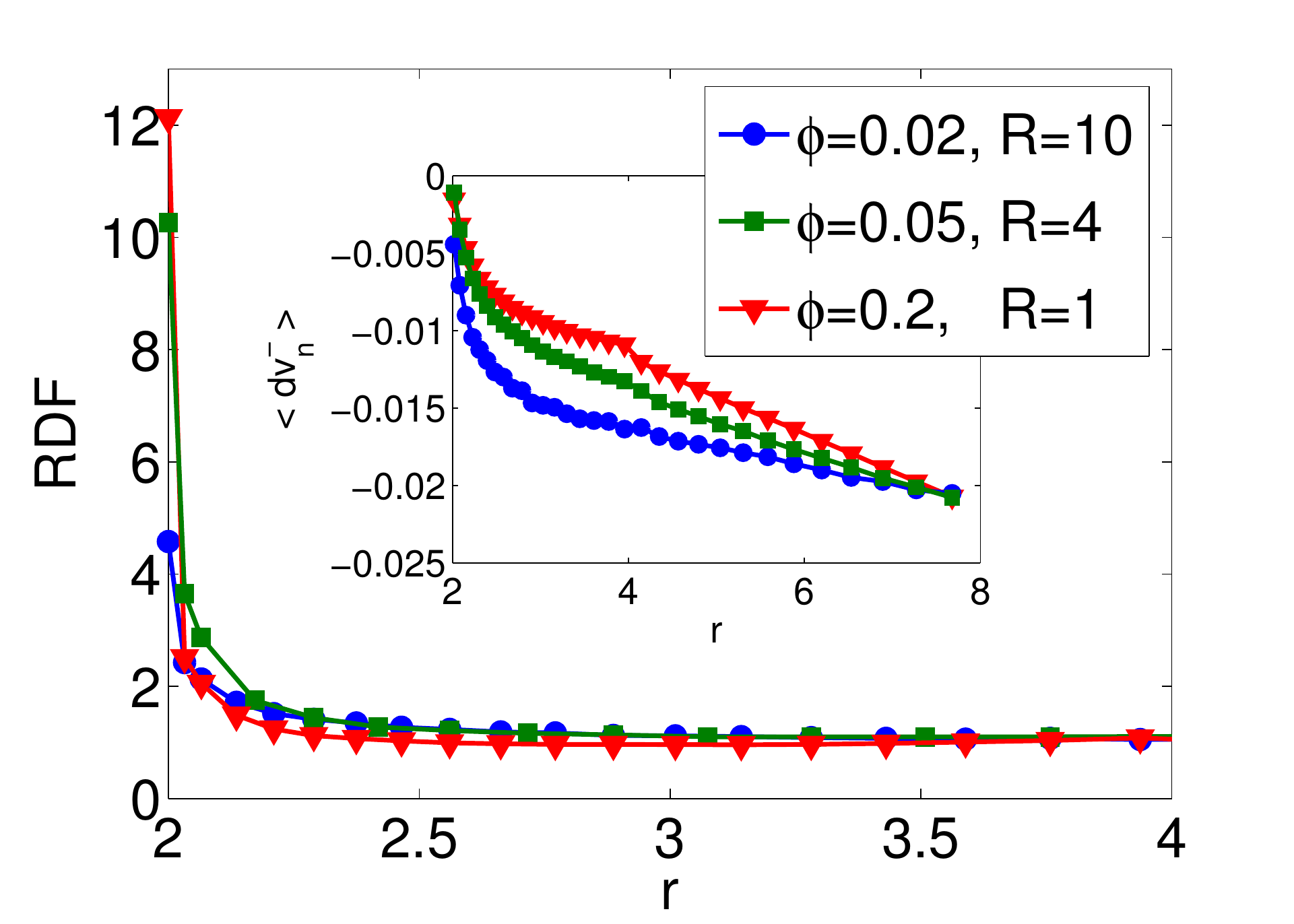}~~~~
\put(-205,110){{\large a)}}
\includegraphics[width=.50\textwidth]{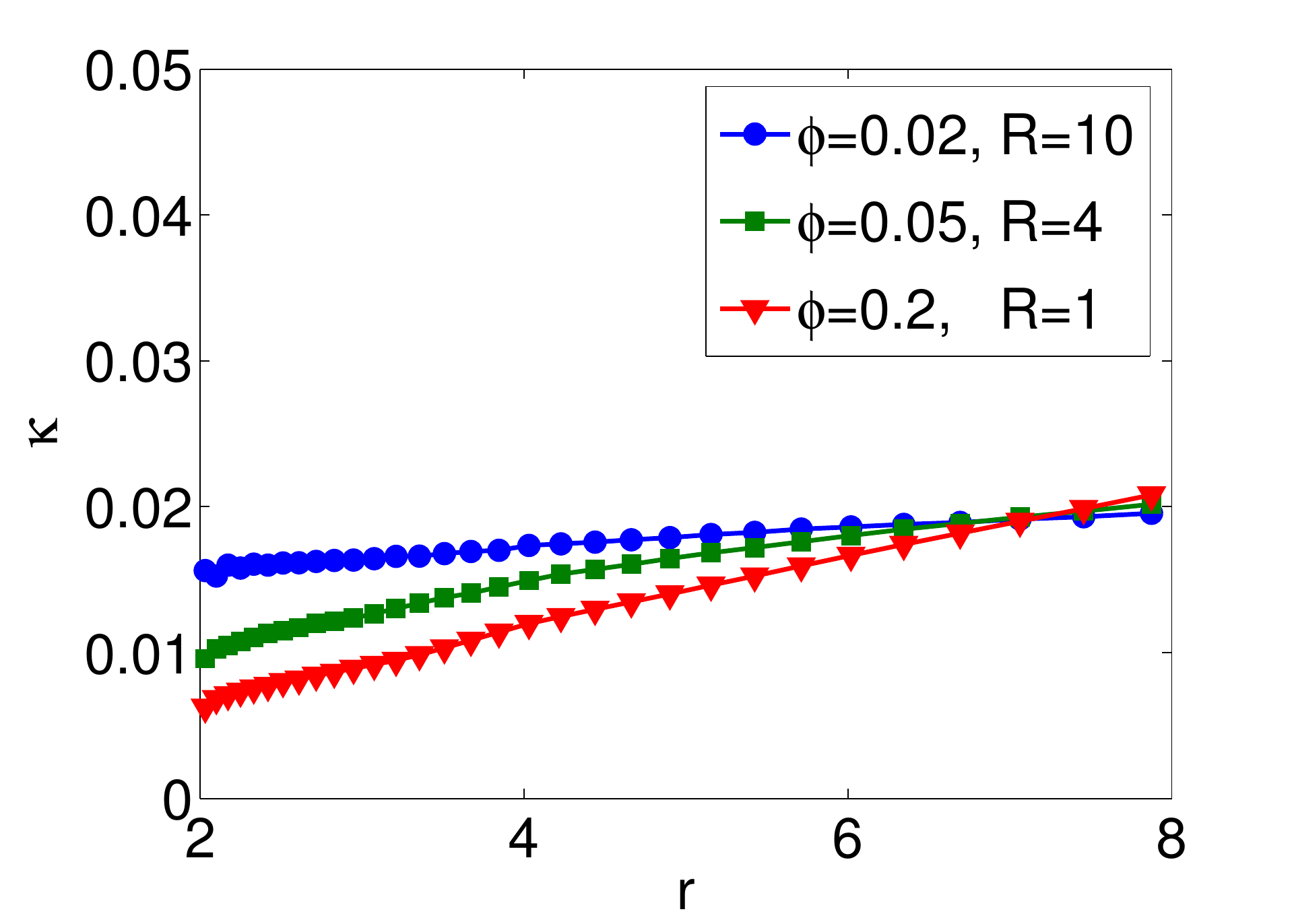}
\put(-205,110){{\large b)}}
\caption{(a) Radial distribution function and average negative relative velocity (inset) and (b) collision kernel (see text for the definitions) around the centerline for the three values of the density ratio $R$ and particle volume fraction $\phi$.}
\label{fig:colli_xi}
\end{figure}

Before concluding the section, we examine the collision statistics when increasing the 
volume fraction $\phi$ while keeping the mass fraction $\chi$ constant. To this aim, we show in figure~\ref{fig:colli_xi} the radial distribution function $RDF$, the
averaged normal relative velocity and the collision kernel from 3 of the cases at constant mass fraction  previously 
discussed: $\phi=2\%$ and $R=10$; $\phi=5\%$ and $R=4$; $\phi=20\%$ and $R=1$.

The small-scale clustering increases as the volume fraction $\phi$ 
increases, see figure~\ref{fig:colli_xi}(a),
 i.e.\ the $RDF$ at $r=2$ is highest for the flow with $\phi=20\%$. 
However, as the excluded volume 
increases with $\phi$, the mean distance between the particles is reduced and these approach each other on average with 
a smaller relative velocity, as shown by the reduction in $\langle dv_n^-(r) \rangle$ 
at higher $\phi$ in the inset of figure~\ref{fig:colli_xi}(a). 
Finally, figure~\ref{fig:colli_xi}(b) reveals that also at constant $\chi$ the collision rate is mainly governed by the averaged normal relative 
velocity. We observe indeed that $\kappa(r)$ is higher in the most dilute cases and the data scale with the volume fraction.

\section{Final remarks}

We study the effect of varying solid to fluid density ratio and volume fraction in a
turbulent channel flow laden with finite-size rigid spheres in the semi-dilute regime.
The numerical simulations do not include the effect of gravity to disentangle the role of fluid and particle inertia, as well as of the excluded volume on the mean and fluctuating fluid velocities and particle motion.

The main finding of the work is that
variations of the volume fraction have a larger impact on the statistics of fluid and solid phases than modifications
of the density ratio $R$.
Indeed, we show that, when the volume fraction is kept constant ($\phi=5\%$) and the density
ratio, $R$, increased from 1 to $R\le10$,
 the mean fluid velocity and velocity fluctuation
profiles are only slightly affected.
The  main effect of increasing the density ratio (up to $R=10$) is the change of the
mean local volume fraction, i.e.\ the wall-normal particle distribution across the channel. At 
$R=10$, we report a significant shear-induced migration toward the centerline.
This is shown to be an inertial effect induced by the particle density, $R$, and the presence of a wall.

When the volume fraction is changed and either the mass fraction or the density ratio kept constant, instead, the flow statistics vary significantly.
The mean streamwise velocity profiles in outer units show lower values closer to the walls and higher values toward the centerline. 
In inner units, the difference is even more evident, showing a continuos variation of the von K\'{a}rm\'{a}n constant and of the
additive coefficient of the log-law, see also Ref.~\onlinecite{picano2015} for comparisons at constant $R=1$. 
The increase in overall drag found when varying the volume fraction is considerably 
higher than that obtained for increasing density ratios at same volume fraction.

We also consider cases at same $\phi=5\%$ and $R=1000$.
At this high $R$, the motion of the solid phase decouples from the dynamics of the fluid phase and the statistics drastically change.
The particles are uniformly distributed across the channel and behave as a dense gas with uniform mean streamwise velocity 
and uniform isotropic velocity fluctuations across the channel.
The dense gas behavior of the solid phase clearly emerges in the probability density function
of the modulus of the velocity fluctuations that closely follows a Maxwell-Boltzmann distribution.
The fluid velocity fluctuations are reduced and are
almost constant except in the regions close to the walls.
For $R=1000$ we also find that the streamwise dispersion is one or two orders of magnitude smaller than in the cases at
lower $R$. In channel flows, the streamwise particle dispersion is enhanced by the inhomogeneity of the mean velocity 
profile. However as we have shown, at very high density ratios this inhomogeneity is lost leading to a reduction of the mean 
streamwise particle displacement.

Finally, we have examined the radial distribution of particles and their collision
kernel.  For $1\le R\le10$ and constant $\phi=5\%$,
the collision rate is mostly controlled by the particle clustering near contact.
Instead, for $R=1000$, the number of collisions is enhanced and essentially determined
by the particle average normal relative velocity.
For suspensions at fixed mass fraction $\chi=0.2$, the collision rate decreases with increasing $\phi$.

Our results therefore suggest that the particle motion in the absence of gravity is not significantly different between neutrally buoyant particles and heavy particles 
with density ratios typical of sediments and metal particles in liquids.
The main effects on the flow statistics are due to variations of the volume fraction, thus of the excluded volume. 
The main effect of increasing the density ratio is the appearance of a shear-induced migration while velocity statistics 
are almost unchanged.
The present results may help to interpret the dynamics of sediments in shear turbulence.

\begin{acknowledgments}
This work was supported by the European Research Council Grant No.\ ERC-2013-CoG-616186, TRITOS, from the Swedish Research Council (VR), through the 
Outstanding Young Researcher Award, and from the COST Action MP1305: \emph{Flowing matter}.
Computer time provided by SNIC (Swedish
National Infrastructure for Computing) and CINECA, Italy (ISCRA Grant FIShnET-HP10CQQF77).
\end{acknowledgments}

\bibliographystyle{aipnum4-1long}
%

\end{document}